\documentclass[11pt,a4paper]{article}
\usepackage{jheppub}
\usepackage{latexsym}
\usepackage{pdflscape}
\usepackage{bm}
\numberwithin{equation}{section}
\allowdisplaybreaks

\allowdisplaybreaks[2]

\def\Gc{\Gamma_\text{cusp}}

\def\cB{\mathcal{B}}
\def\cS{\mathcal{S}}

\def\cO{\mathcal{O}}

\def\cN{\mathcal{N}}
\def\cR{\mathcal{R}}
\def\cK{\mathcal{K}}
\def\mint{\int_{-\infty}^\infty\!\cdots\!\int_{-\infty}^\infty}

\newcommand{\be}{\begin{equation}}
\newcommand{\ee}{\end{equation}}
\newcommand{\ba}{\begin{aligned}}
\newcommand{\ea}{\end{aligned}}

\def\({\left(}
\def\){\right)}

\DeclareMathOperator{\real}{Re}
\DeclareMathOperator{\im}{Im}

\DeclareMathOperator{\Disc}{Disc}

\newcommand{\re}{{\rm e}}
\newcommand{\ri}{{\rm i}}
\newcommand{\rd}{{\rm d}}

\title{Resurgence of the Cusp Anomalous Dimension}

\author{Daniele Dorigoni}
\author{and Yasuyuki Hatsuda}

\affiliation{DESY Theory Group, DESY Hamburg, \\
Notkestrasse 85, D-22603 Hamburg, Germany}
\emailAdd{daniele.dorigoni@desy.de} 
\emailAdd{yasuyuki.hatsuda@desy.de}

\abstract{
We revisit the strong coupling limit of the cusp anomalous dimension in
planar $\mathcal{N}=4$ super Yang-Mills theory. It is known that the strong coupling expansion
is asymptotic and non-Borel summable. As a consequence, the cusp anomalous dimension
receives non-perturbative corrections, and
the complete strong coupling expansion should be a resurgent transseries.
We reveal that the perturbative and non-perturbative parts in the transseries
are closely interrelated.
Solving the Beisert-Eden-Staudacher equation systematically, we analyze in detail the large order behavior in
the strong coupling perturbative expansion and show that the non-perturbative information is 
indeed encoded there.
An ambiguity of (lateral) Borel resummations of the perturbative expansion is precisely canceled by the contributions 
from the non-perturbative sectors, 
and the final result is real and unambiguous.
}

\begin{document}

\maketitle

\renewcommand{\thefootnote}{\arabic{footnote}}
\setcounter{footnote}{0}
\setcounter{section}{0}

\section{Introduction}\label{sec:intro}

In recent years, there is remarkable progress in the understanding of the AdS/CFT duality \cite{Maldacena, GKP1, Witten}.
In planar AdS/CFT between $\cN=4$ super Yang-Mills (SYM) theory and string theory
on $AdS_5 \times S^5$, \textit{integrability} is a key concept in the investigation.
Here we refer to \cite{Integrability} for  a comprehensive review on the AdS/CFT integrability,
but of course there are many important results after that review.
In particular, recently a new formulation, called Quantum Spectral Curve, to solve the spectral problem
in planar $\cN=4$ SYM was proposed in \cite{GKLV1, GKLV2}.

In this work, we focus on a well-studied quantity called the cusp anomalous dimension $\Gc(g)$ (or sometimes called the scaling
function).
Throughout this paper, we use the conventional notation of the coupling constant $g$, 
which is related to the 't Hooft coupling $\lambda=g_\text{YM}^2 N$ by
\be
g=\frac{\sqrt{\lambda}}{4\pi}.
\ee
The cusp anomalous dimension appears in several contexts.
It appears as a UV divergence of a (light-like) Wilson operator with a cusp or as
an IR divergence of a gluon scattering amplitude.
These divergences are closely related by the so-called Wilson loop/amplitude duality \cite{AM1, Drummond:2007aua, Brandhuber:2007yx, Drummond:2007au}.
It is well-known that the cusp anomalous dimension also appears as a logarithmic divergence
of conformal dimension $\Delta$ for a twist-two operator with large spin $S$:
\be
\Delta-S=2\Gc(g) \log S+\cdots,\qquad S \to \infty.
\label{eq:Delta}
\ee
Note that this logarithmic behavior is universal for any $g$,
thus the cusp anomalous dimension is a good interpolating function from weak to strong coupling.
From these examples, it is obvious that to understand the cusp anomalous dimension is an important task.

Very surprisingly, the cusp anomalous dimension in planar $\cN=4$ SYM can be computed at
\textit{any coupling}
by solving the so-called Beisert-Eden-Staudacher (BES) equation \cite{BES} (see also \cite{ES})!
This is one of the greatest achievements in the AdS/CFT integrability.
We can learn many things through this equation.
At weak coupling, the cusp anomalous dimension admits the standard perturbative expansion in $g^2$:
\be
\Gc(g)=4g^2\left[ 1-\frac{\pi^2g^2}{3}+\frac{11\pi^4g^4}{45}-2\( \frac{73\pi^6}{630}-4\zeta(3)^2 \) g^6+\cO(g^{8}) \right],
\quad g \to 0.
\label{eq:cusp-weak}
\ee
As observed in \cite{BES}, this weak coupling expansion is a \textit{convergent} series with finite radius $g=1/4$.
Therefore, it is expected that there are no non-perturbative corrections%
\footnote{To be precise, a convergent series is not a sufficient condition for non-existence of 
non-perturbative corrections. A counter-example is the exact planar free energy in ABJM theory \cite{ABJM}.
As shown in \cite{DMP1}, it receives a non-perturbative correction of the form $\re^{-2\pi \sqrt{2\lambda}}$ at strong coupling
even though its perturbative $1/\sqrt{\lambda}$ expansion is convergent. 
The same also happens for the so-called interpolation function \cite{Gromov:2014eha} (see \eqref{eq:h-strong}).}
of the form $\re^{-A/g}$, and the perturbative expansion \eqref{eq:cusp-weak} 
is sufficient to reconstruct the full function $\Gc(g)$.
The weak coupling result was confirmed up to four loops \cite{Bern:2006ew, Cachazo:2006az}.

At strong coupling, the problem is much more involved.
The BES equation predicts the strong coupling expansion
\be
\Gc(g)=2g \( 1-\frac{3 \log 2}{4\pi g}-\frac{\mathrm{K}}{16\pi^2 g^2}
+\cdots\), \qquad g \to \infty, 
\label{eq:cusp-strong}
\ee
where $\mathrm{K}=0.915965594\dots$ is Catalan's constant.
These coefficients were first predicted in \cite{BBKS} by the numerical analysis.
In \cite{AABEK, Kostov:2007kx, Beccaria:2007tk}, the leading coefficient was computed analytically, and then,
in \cite{BKK}, the expansion was systematically computed up to $1/g^{40}$.
Of course, the strong coupling prediction \eqref{eq:cusp-strong} should be compared with the direct string worldsheet computation.
In fact, the worldsheet computation up to two loops shows that the strong coupling prediction \eqref{eq:cusp-strong} 
is perfectly reproduced \cite{GKP2, Kruczenski:2007cy, Roiban:2007jf, Roiban:2007dq}.
This agreement is obviously a strong evidence of the planar AdS/CFT duality.

However, this is not the end of the story at strong coupling.
A crucial observation in \cite{BKK} is that the strong coupling perturbative expansion \eqref{eq:cusp-strong}
is very likely \textit{asymptotic} and \textit{non-Borel summable} due to singularities
on the positive real axis in Borel plane.
At first glance, the non-Borel summability causes a serious problem
in the perturbative resummation, and it seems that the Borel resummation procedure
does not work any more. 
However, there is a beautiful resolution for this resummation problem.
As well-known, the Borel singularities on the positive real axis cause an ambiguity of Borel resummations
because one has to avoid the singularities when performing the inverse Borel transform (see figure~\ref{Fig:Stokes}).
There are several choices for the integration contours to avoid the singularities.
The important point is that the ambiguity has non-perturbative order $\re^{-A g}$,
where $A$ is related to the closest Borel singularity to the origin.
Therefore a natural expectation is that there is a non-perturbative correction of the same order,
and that its Borel resummation precisely cancels
the ambiguity in the perturbative resummation.
As a result, the total sum should give the same answer for any choice of the contours in the Borel resummations. 

This beautiful structure of cancellation of ambiguities goes under the name of resurgence and it was developed by Ecalle \cite{Ecalle:1981}. Since then it has been applied at first in quantum mechanical systems \cite{ZinnJustin:1980uk,Voros:1983} and only very recently it has been applied to quantum field theory \cite{Dunne:2012zk,Argyres:2012vv,Cherman:2013yfa} to obtain a weak coupling interpretation of the IR renormalons .

Similarly, in the cusp anomalous dimension, 
the Borel resummation of the strong coupling perturbative expansion alone is \textit{insufficient} to reconstruct the full function $\Gc(g)$.
To resolve the Borel resummation problem, we need non-perturbative corrections.
In a series of works \cite{BKK, BK1, BBBKP, BK2}, it turned out that the cusp anomalous dimension should receives the following non-perturbative corrections
\be
\ba
\frac{\Gc(g)}{2g}=\Gc^{(0)}(g)-\frac{\Lambda^2}{2\pi g} \Gc^{(1)}(g)+\frac{ \Lambda^4}{16\pi^2 g^2} \Gc^{(2)}(g)+\cO(\Lambda^6), \qquad g \to \infty,
\ea
\label{eq:cusp-trans}
\ee
where $\Gc^{(0)}(g)$ is the perturbative contribution above and $\Lambda^2$ is a non-perturbative scale, related to the 't Hooft coupling as follows
\be
\Lambda^2=\sigma \frac{\Gamma(\frac{3}{4})}{\Gamma(\frac{5}{4})}(2\pi g)^{1/2} \re^{-2\pi g}. 
\label{eq:Lambda}
\ee
Note that the (complex) parameter $\sigma$ depends on a choice of Borel resummations, as will be seen later.
The non-perturbative scale is closely related to the mass gap of the O(6) sigma model\footnote{The O(6) sigma model, being an asymptotically free two-dimensional field theory, is affected by the so called IR renormalons. See \cite{Dunne:2015ywa} for a recent discussion of the resurgence properties of the O(N) models and the connections with the IR renormalons problem.},
which describes the effective string worldsheet theory in the high spin limit, as explained in \cite{AM2} (see also \cite{Basso:2010in}). 
In each non-perturbative sector, $\Gc^{(n)}(g)$ has the asymptotic (non-Borel summable) $1/g$ expansion%
\footnote{One can always set $ \Gamma_0^{(n)}=1$ by factoring out an appropriate factor.}:
\be
\\
\Gc^{(n)}(g)=\sum_{\ell=0}^\infty \frac{\Gamma_\ell^{(n)}}{(2\pi g)^{\ell}}.
\ee
In principle, one can compute the coefficients $\Gamma_\ell^{(n)}$ by solving the BES equation at strong coupling,
but this is a highly non-trivial task.
A very first few coefficients in the leading non-perturbative sector $\Gc^{(1)}(g)$ were computed in \cite{BK2}.

One of our goals in this paper is to show that there are non-trivial relations among the coefficients $\Gamma_\ell^{(n)}$.
In particular, the perturbative coefficients $\Gamma_\ell^{(0)}$ already know information
on the non-perturbative sector!
See \eqref{eq:LOB-pert} for a concrete result.
This is clearly a consequence of the resurgent analysis \cite{Dorigoni:2014hea},
recently developed in many contexts in theoretical physics.
To confirm this fact, we develop the method in \cite{BK2}, and
numerically compute the coefficients $\Gamma_\ell^{(n)}$ ($n=0,1,2$) up to $\ell=180$
with sufficiently high precision.
The fact that the perturbative and non-perturbative parts are interrelated to each other provides us 
a strong consistency
test of the strong coupling solution.
If an obtained solution is wrong, one would encounter a discrepancy with the resurgent expectation.
Moreover, one can predict the non-perturbative correction from the perturbative result
\textit{without solving the BES equation}.
We also demonstrate that the ambiguity of the Borel resummations in the perturbative sector
is precisely canceled by the same ambiguity in the non-perturbative sector.
As a consequence, the final answer is always real-valued and unambiguous.
Our results show that the cusp anomalous dimension at strong coupling is a resurgent transseries.

The organization of this paper is as follows.
In Section \ref{sec:Borel},  we briefly review some basic concepts in Borel resummation and resurgence theory, we refer to \cite{Dorigoni:2014hea} for a longer discussion. We discuss then, in Section \ref{sec:BES}, how to obtain the strong coupling expansion of the cusp anomalous dimension from the solution to the BES equation. To better understand the resurgent properties of the cusp anomaly in $\mathcal{N}=4$, in Section \ref{sec:toy} we analyze a toy model solution to the BES equation previously presented in \cite{BK2}. In Section \ref{sec:cusp}, we finally study the cusp anomaly in $\mathcal{N}=4$ and show how to reconstruct all the non-perturbative corrections, i.e. the full transseries solution for $\Gc(g)$, simply from the perturbative strong coupling asymptotic expansion. We show how the transseries representation for $\Gamma_\text{cusp}(g)$ is free from ambiguities and we comment on the relation between the cusp anomaly and the mass gap of the O(6) sigma model.
We draw our conclusions in \ref{sec:conc}.
 
 \vspace{0.2cm}

\paragraph{Note added:} While this work was being completed, we became aware of
related work \cite{Aniceto:2015rua}, which has some overlap with this paper.

\section{Borel resummations and resurgent transseries}\label{sec:Borel}

As we will review in details later on, in the strong coupling regime the perturbative contribution to the cusp anomaly, $\Gc^{(0)}(g)$, and the perturbative corrections in each non-perturbative sector, $\Gc^{(n)}(g)$, all take the form of asymptotic power series in $1/z$ ($z=2\pi g$)
\be
f(z) =c + \sum_{n=0}^\infty \frac{f_n}{z^{n+1}},\qquad z \to \infty \,,\label{eq:PowerSeries}
\ee
where the coefficients $f_n$ diverge factorially like $\Gamma(n+\alpha)$, for some constant $\alpha$, denoting a Gevrey-1 type series.
For the terminology in the Borel analysis, see \cite{Dorigoni:2014hea}, for instance.

For this kind of asymptotic power series, the standard Borel transform\footnote{If the coefficients $f_n$ grows asymptotically as $\Gamma(n+\alpha)$, it is sometimes more useful to use a slightly different Borel transform where the coefficient $f_n$ gets divided precisely by $\Gamma(n+\alpha)$.} proves to be an extremely useful object, which is defined as
\be
\mathcal{B}[f](t) = \sum_{n=0}^\infty \frac{f_n}{\Gamma(n+1)} t^n\,.
\ee 
Note that this series converges at least in a neighborhood of the origin $t=0$.
We can thus obtain an analytic continuation of the original formal power series by a Laplace transform of $B[f]$
\be
\mathcal{S}_\theta f(z)  = c+\int_0^{\re^{\ri \theta} \infty} \rd t\,\re^{-t \,z}\, \mathcal{B}[f](t)\,,\label{eq:StdBorel}
\ee
usually called Borel resummation of $f(z)$ in the direction $\theta$.

When $\mathcal{B}[f](t)$ contains singularities along a particular direction $\theta$, in the complex $t$-plane, also called Borel plane, we will say that $\theta$ is a Stokes line for $f$ (or equivalently $\mathcal{B}[f]$). Along a Stokes line we cannot directly use the resummation $\mathcal{S}_\theta$, but we can easily dodge the singularities by defining the two lateral resummations
\be
\ba
\mathcal{S}_{\theta^+} f(z)&= c+\int_0^{\re^{ \ri\,(\theta+\epsilon)}\infty} \rd t\, \re^{-t\,z} \,\mathcal{B}[f](t)\,,\\
\mathcal{S}_{\theta^-} f(z)&= c+\int_0^{\re^{ \ri\,(\theta-\epsilon)}\infty} \rd t\,\re^{-t\,z} \,\mathcal{B}[f](t)\,.
\ea
\label{eq:lateral}
\ee
Note that if the direction $\theta$ does not contain any singularity, the two lateral summations coincide with each others and with the standard Borel resummation (\ref{eq:StdBorel}): $\mathcal{S}_{\theta}=\mathcal{S}_{\theta^+}=\mathcal{S}_{\theta^-}$.

When $\theta$ is a Stokes line, we have $\mathcal{S}_{\theta^+}\neq\mathcal{S}_{\theta^-}$, but it is still possible to relate the two different analytic continuations of the asymptotic series $f(z)$ via the so-called Stokes automorphism, $\mathfrak{S}_\theta$, in the direction $\theta$
\be
\ba
&\mathcal{S}_{\theta^+} = \mathcal{S}_{\theta^-} \circ \mathfrak{S}_\theta = \mathcal{S}_{\theta^-}\circ \left(\mbox{Id}-\mbox{Disc}_\theta\right)\label{eq:Stokes}\,,\\
&\mathcal{S}_{\theta^+}-\mathcal{S}_{\theta^-}=-\mathcal{S}_{\theta^-}\circ \mbox{Disc}_\theta\,.
\ea
\ee
Where $\mbox{Disc}_\theta$ encodes the full discontinuity across $\theta$.

\begin{figure}
\begin{center}
	\includegraphics[scale=0.45]{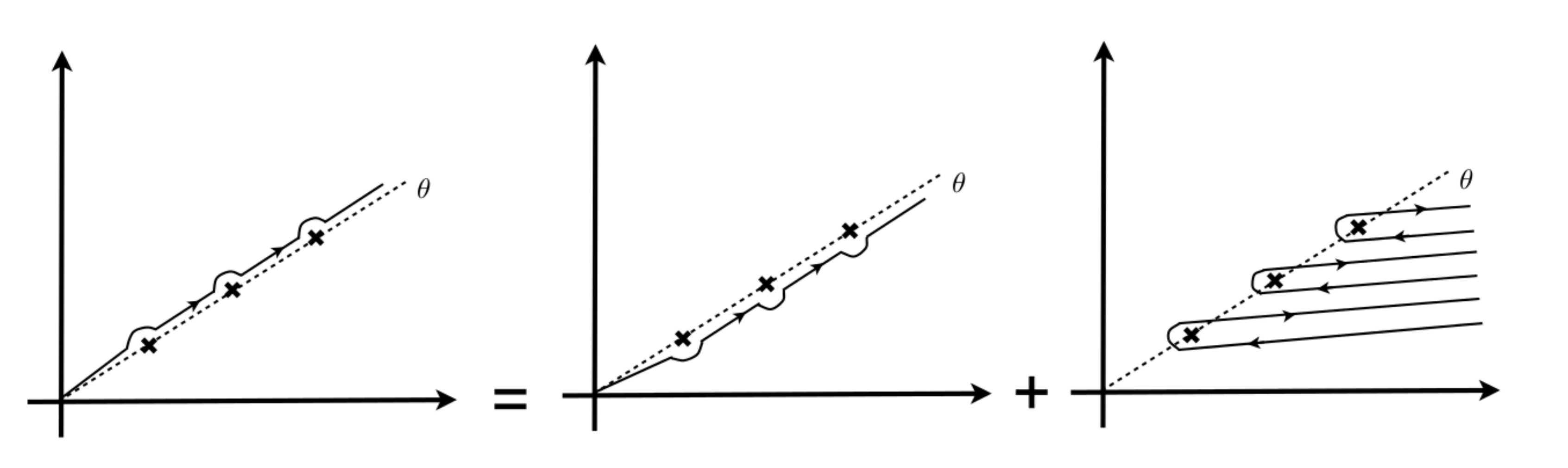}
	\caption{The difference between left and right resummation along the singular direction $\theta$ as a sum over Hankel contours.}
	\label{Fig:Stokes}
\end{center}
\end{figure}

By a simple contour deformation, we can rewrite the difference between the two resummations along $\theta^+$ and $\theta^-$ as a sum over Hankel's contours, and the discontinuity of $\mathcal{S}$ across $\theta$ is given as an infinite sum of contribution coming from each one of the singular points, see Figure \ref{Fig:Stokes}.

From simple residue theory, it is easy to see that the difference between the two resummations generates a new type of non-analytic term, beyond the realm of formal power series, schematically of the form
\be
(\mathcal{S}_{\theta^+}-\mathcal{S}_{\theta^-})\sim 2\pi \ri\, \re^{-t_\ast\,z} \left(a+O(1/z)\right)\,,
\ee
where the point $t_\ast$ is a singularity of $\mathcal{B}[f](t)$ in the direction $\theta$, i.e., $\arg(t_\ast)=\theta$.
These exponentially suppressed terms are nothing new, in semi-classical calculations we know that generically the perturbative expansion receives non-perturbative corrections usually referred to as instantons corrections.
So, whenever we have to deal with an asymptotic perturbative expansion, we expect that the most general solutions to our problem, in the case at hand the cusp anomalous dimension at strong coupling, will take the \textit{transseries} form
\be
F(z) = \sigma_0 \,f^{(0)}(z)+ \sigma_1 \, \re^{-t_1\,z} f^{(1)}(z) + \sigma_2 \, \re^{-t_2\,z} f^{(2)}(z) + ...\,,
\ee
where the exponentially suppressed terms $e^{-t_i\,z}$ correspond to non-perturbative corrections with (possibly complex)``instanton actions'' $t_i$, while the asymptotic power series $f^{(i)}(z)\sim\sum_{n\geq 0} f^{(i)}_n z^{-n-1}$ correspond to the perturbative expansions on top of the non-perturbative corrections\footnote{Despite all of the above series being only asymptotic and the transseries expansion being a formal coding of the exact function, there is a precise, algorithmic way to compute numerically all the coefficients $f^{(i)}_n$ once we are given the exact function $F(g)$ with arbitrary numerical precision. We thank Slava Rychkov for discussions on this problem.}.
The complex numbers $\sigma_i$ are called transseries parameters, they are constants only in wedges of the complex plane, $\theta_1<\arg(z)<\theta_2$,  but they can jump precisely when $\arg(z)$ crosses a Stokes line.

As we will see in a concrete example in Section \ref{sec:toy}, from the sum over Hankel's contours as in Figure \ref{Fig:Stokes}, we can compute the discontinuity of the analytic continuation of the perturbative expansion $f^{(0)}(z)$ and this is given by
\be
(\mathcal{S}_{\theta^+}-\mathcal{S}_{\theta^-} ) f^{(0)}(z) = A_1 \,\re^{- t_1 \,z} f^{(1)}(z) + A_2\,\re^{-t_2\,z} f^{(2)}(z) +...\label{eq:Disc}
\ee
where the only $t_i$ appearing are the one corresponding to all the singular points in the direction $\theta$, i.e. $\arg t_i=\theta$, the constants $A_i$ are complex numbers usually called Stokes constants (or analytic invariants) while the $f^{(i)}(z)$ are precisely the asymptotic formal power series in $1/z$ associated to the non-perturbative corrections $\re^{- t_i \,z} $. 

In a similar way, the asymptotic series $f^{(i)}(z)$ will have singular directions with non-trivial Stokes automorphisms that will involve the other expansions $f^{(j)}(z)$. In this way the ambiguities in resumming separately one by one each asymptotic series $f^{(i)}(z)$ combines together with the jump in the transseries parameters, and the transseries expansion $F(z)$ is an exact representation of the exact answer via Borel-Ecalle resummation
\be
\ba
\mathcal{S}_\theta F(z)& = \sigma^{(+)}_0 \,\mathcal{S}_{\theta^+}f^{(0)}(z)+ \sigma^{(+)}_1 \, \re^{-t_1\,z} \mathcal{S}_{\theta^+} f^{(1)}(z) + \sigma^{(+)}_2 \, \re^{-t_2\,z} \mathcal{S}_{\theta^+}f^{(2)}(z) + \cdots\\
&= \sigma^{(-)}_0 \,\mathcal{S}_{\theta^-}f^{(0)}(z)+ \sigma^{(-)}_1 \, \re^{-t_1\,z} \mathcal{S}_{\theta^-} f^{(1)}(z) + \sigma^{(-)}_2 \, \re^{-t_2\,z} \mathcal{S}_{\theta^-}f^{(2)}(z) + \cdots\,,
\ea
\ee
where the complex numbers $\sigma^{(\pm)}_i$ are the transseries parameters in the respective wedges $\arg(z)> \theta+\epsilon$, $\arg(z) <\theta -\epsilon$.
Even if we do not have at our disposal the full transseries, we can still obtain a great deal of information simply by studying the discontinuity properties, across a Stokes line, of the resummation of $f^{(0)}(z)$, or similarly the large orders behavior of the perturbative coefficients $f^{(0)}_n$.

Let's suppose that the only Stokes line for $f^{(0)}(z)$ is the positive real axis, $\theta=0$, then by Cauchy theorem\footnote{Note that generically we would get contributions coming from all the discontinuities across every singular directions in the Borel plane.} \cite{ZinnJustin:1980uk} we know that
\begin{equation}
f^{(0)}(z) = \frac{1}{2\pi \ri} \oint \! \rd \omega \frac{f^{(0)}(\omega)}{\omega-z}
=\frac{1}{2\pi \ri} \int_0^{\infty} \rd \omega \frac{\Disc_0 f^{(0)}(\omega)}{\omega-z} \,,
\end{equation}
where we dropped the contribution coming from infinity,
and by expanding for $z \to \infty$
\begin{equation*}
\frac{1}{\omega-z} = - \sum_{n=0}^\infty \omega^n\,z^{-n-1}\,,
\end{equation*}
we get
\begin{equation}
f^{(0)}_n \sim -\frac{1}{2\pi \ri} \int_0^\infty \rd\omega\,\omega^{n} \Disc_0 f(\omega) \,,\label{eq:Largeorder}
\end{equation}
where we schematically wrote $f^{(0)}(z) \sim \sum_{n\geq 0} f_n z^{-n-1}$.

We can now use (\ref{eq:Disc}) to relate the large order coefficients $f^{(0)}_n$ to the lower order coefficients of the asymptotic expansion $f^{(i)}(z)\sim\sum_{n\geq 0} f^{(i)}_n z^{-n-1}$ (we refer to the thorough work of Aniceto, Schiappa and Vonk \cite{Aniceto:2011nu} for a more general discussion), then
\be
f^{(0)}_n \sim \frac{1}{2\pi \ri} \int_0^\infty \rd\omega\,\omega^{n} \left[ A_1 \re^{-t_1\,\omega} \left(  \frac{f_0^{(1)}}{\omega} +\frac{ f_1^{(1)} }{\omega^2}+\cdots  \right) +A_2 \re^{-t_2\,\omega} \left(  \frac{f_0^{(2)}}{\omega}+\cdots \right)+\cdots\right]\,,
\ee
which leads to
\begin{align}
f^{(0)}_n \sim&\notag \frac{A_{1}}{2\pi \ri}\frac{n!}{(t_1)^{n+1}} \left( f_0^{(1)} + f_1^{(1)} \frac{t_1}{n} + f_2^{(1)} \frac{t_1^2}{n(n-1)}+f_3^{(1)} \frac{t_1^3}{n(n-1)(n-2)}+\cdots\right)\\
&+\frac{A_{2}}{2\pi \ri}\frac{n!}{(t_2)^{n+1}} \left( f_0^{(2)} + f_1^{(1)} \frac{t_2}{n} + f_2^{(2)} \frac{t_1^2}{n(n-1)}+\cdots\right)
+\cdots\,. \label{eq:HigherOrderP}
\end{align}
This beautiful result tells us that the large order coefficients of the perturbative expansion do contain explicitly the lower order coefficients of the non-perturbative contributions expansion (and vice versa)!
Furthermore we want to stress that the large order behavior of the perturbative expansion can also be use to determine whether the problem at hand is resurgent or not \cite{berry1993unfolding}. 
Had we forgotten to include some non-perturbative contribution \cite{BalianPV,Couso-Santamaria:2014iia,Vonk:2015sia}, or some transseries parameter (see Section \ref{sec:cusp}), we would have found that the large orders of the perturbative expansion do not satisfy (\ref{eq:HigherOrderP}). We can ``experimentally" check that, to obtain the correct asymptotic behavior for the perturbative coefficients $f_n$, new additional transseries parameters and/or new instanton types have to be added to our original transseries ansatz!

\section{Solving the BES equation}\label{sec:BES}

The starting point of our analysis is the BES equation \cite{BES}.
In this section, we review how to solve this equation at strong coupling.
In particular, we would like to compute the strong coupling expansion as high as possible.
For this purpose, we follow the approach in \cite{BKK, BK2} (see also \cite{Kostov:2008ax, Volin}).

The BES equation is an integral equation for a rapidity density of a twist-two operator
with large spin.
It is derived from the asymptotic Bethe ansatz equations.
The explicit form of the BES equation is given by
\be
\sigma(t)=\frac{t}{\re^t-1} \left[ K(2gt,0)-4g^2 \int_0^\infty \rd t' \, K(2gt, 2gt') \sigma(t') \right].
\label{eq:BES}
\ee
where the kernel $K(t,t')$ consists of two parts: the ``main scattering'' part and the ``dressing phase'' part:
\be
\ba
K(t,t')&=K_\text{m}(t,t')+K_\text{d}(t,t'), \\
K_\text{m}(t,t')&=K_0(t,t')+K_1(t,t'),\\
K_\text{d}(t,t')&=8g^2 \int_0^\infty \rd t'' \, K_1(t, 2g t'')\frac{t''}{\re^{t''}-1} K_0(2gt'',t'),
\ea
\ee
with
\be
\ba
K_0(t,t')=\frac{t J_1(t)J_0(t')-t' J_0(t)J_1(t')}{t^2-{t'}^2}, \quad
K_1(t,t')=\frac{t' J_1(t)J_0(t')-t J_0(t)J_1(t')}{t^2-{t'}^2}.
\ea
\ee
The unknown function $\sigma(t)$ is the density to be solved.
Once this function is known, the cusp anomalous dimension is given by
\be
\Gc(g)=8g^2 \sigma(0).
\ee
Therefore our task is to solve the BES equation \eqref{eq:BES}.
We stress that the BES equation is believed to be valid at any coupling.
It smoothly interpolates between the weak coupling regime and the strong coupling regime.
At weak coupling, it is not hard to solve \eqref{eq:BES} directly order by order in $g^2$ \cite{BES},%
\footnote{However, it is more convenient to use a discrete matrix formulation in subsection~\ref{subsec:numerical} \cite{BBKS}
in the systematic higher order computation.}
while to find the strong coupling solution is not straightforward.
A direct approach to solve the BES equation at strong coupling (and at finite coupling numerically) is
found in \cite{BBKS, AABEK}.
Below, we review the method developed in \cite{BKK, BK2}.
This approach is indeed very powerful to find the strong coupling solution systematically.

\subsection{Strong coupling solution}
To find the strong coupling solution, we first divide $\sigma(t)$ into even/odd parity parts:
\be
\frac{\re^t-1}{t} \sigma(t)=\frac{\gamma_+(2gt)+\gamma_-(2gt)}{2gt},
\qquad \gamma_\pm(-t)=\pm \gamma_\pm (t).
\label{eq:gamma-pm}
\ee
Then the BES equation leads to the following equations:
\be
\ba
&\int_0^\infty \frac{\rd t}{t} \left[ \frac{\gamma_-(2gt)}{1-\re^{-t}}+\frac{\gamma_+(2gt)}{\re^{t}-1} \right]J_{2n-1}(2gt)
=\frac{1}{2}\delta_{n1}, \\
&\int_0^\infty \frac{\rd t}{t} \left[ \frac{\gamma_+(2gt)}{1-\re^{-t}}-\frac{\gamma_-(2gt)}{\re^{t}-1} \right]J_{2n}(2gt)
=0.
\ea
\label{eq:BES-rewrite0}
\ee
Next we introduce two functions $\Gamma_\pm (t)$ by
\be
\widehat{\Gamma}(t) \equiv \Gamma_+(t) +\ri \Gamma_-(t)=\( 1+\ri \coth \frac{t}{4g} \)(\gamma_+(t)+\ri \gamma_-(t)).
\label{eq:Gamma-def}
\ee
As was shown in \cite{BK2}, the equations \eqref{eq:BES-rewrite0} is rewritten as the form
\be
\int_0^\infty \rd t\, [ \re^{\ri u t} \Gamma_-(t)-\re^{-\ri u t} \Gamma_+(t) ]=2,\qquad
(|u|<1).
\label{eq:BES-rewrite}
\ee
The important point is that the solution $\widehat{\Gamma}(t)$ must have an infinite number of zeros at $t=4\pi \ri  g (m-1/4)$ ($m \in \mathbb{Z}$)
and poles at $t=4\pi \ri g m'$ ($m' \in \mathbb{Z}\backslash \{0\}$).
This is a reflection of the factor $1+\ri \coth \frac{t}{4g}$ in \eqref{eq:Gamma-def} and the entireness of $\sigma(t)$.
This analyticity condition uniquely fix the solution of the equation \eqref{eq:BES-rewrite}.
In general, the equation \eqref{eq:BES-rewrite} alone does not fix its solution uniquely.
According to \cite{BK2}, a solution of \eqref{eq:BES-rewrite} without an analyticity condition 
is generically given by
\be
\widehat{\Gamma}(\ri t)=f_0(t)V_0(t)+f_1(t) V_1(t),\label{eq:Gammahat}
\ee
where $V_0(t)$ and $V_1(t)$ are known functions, whose explicit forms are presented in appendix~\ref{sec:special} (see \eqref{eq:Vn-def}).
The unknown functions $f_0(t)$ and $f_1(t)$ are fixed after requiring the analyticity condition.
The cusp anomalous dimension is then given by
\be
\Gc(g)=-2g \widehat{\Gamma}(0)=2g(1-2f_1(0)),
\label{eq:Gamma-cusp1}
\ee
where we have used $V_0(0)=1$, $V_1(0)=2$ and the non-trivial equation (see \cite{BK2})
\be
f_0(0)=-1.
\label{eq:f0-t0}
\ee
The condition that $\widehat{\Gamma}(t)$ has the zeros at $t=4\pi \ri  g (m-1/4)$ ($m \in \mathbb{Z}$) leads to
a quantization condition
\be
f_0(t_m)V_0(t_m)+f_1(t_m)V_1(t_m)=0,\qquad t_m=4\pi g \(m -\frac{1}{4} \).
\label{eq:QC}
\ee
For later convenience, we rewrite it as
\be
f_0(4\pi g x_m)+r(x_m)f_1(4\pi g x_m)=0,\qquad x_m=m-\frac{1}{4},
\label{eq:QC2}
\ee
where
\be
r(x)=\frac{V_1(4\pi g x)}{V_0(4\pi g x)}.
\label{eq:r}
\ee
Note that this quantization condition is valid at arbitrary coupling.
In \cite{BK2}, the strong coupling solution of the quantization condition \eqref{eq:QC2} was constructed.
One important remark is that the asymptotic behavior of $r(x)$ at strong coupling is different
for $x>0$ and $x<0$ (and also for $\real g >0$ and $\real g<0$).
This is clearly understood by looking at the relation \eqref{eq:V-U}.
It is also important to note that $r(x_m)$ has a non-perturbative correction of order
\be
r^\text{np}(x_m) \sim \cO(\Lambda^{|8m-2|}),\qquad m \in \mathbb{Z}.
\ee
See \eqref{eq:r-strong-1} and \eqref{eq:r-strong-2} for full detail.
Therefore, up to $\cO(\Lambda^4)$, only $r^\text{np}(x_0)$ contributes to the quantization condition,
while at $\cO(\Lambda^6)$, $r^\text{np}(x_1)$ also appears.
In table \ref{tab:r}, we show for small values of $m$, the order in the non-perturbative scale $\Lambda$, for which $r(x_m)$ is going to contribute.

\begin{table}
  \caption{\label{tab:r}First few non-perturbative contributions present in $r(x_m)$ for different values of $m\in\mathbb{Z}$.}
\begin{center}
  \begin{tabular}{  | c | c || c || c | c | }
    \hline
    $-2$ & $-1$ & $m=0$ & $+1$ & $+2$ \\ \hline\hline
   $ \Lambda^0 $&$ \Lambda^0 $&$ \Lambda^0 $&$ \Lambda^0 $&$ \Lambda^0 $\\  \hline
   $ \Lambda^{18} $&$ \Lambda^{10} $&$ \Lambda^2 $&$ \Lambda^6 $&$ \Lambda^{14} $\\
    \hline
       $ \Lambda^{36} $&$ \Lambda^{20} $&$ \Lambda^4 $&$ \Lambda^{12} $&$ \Lambda^{28} $\\
    \hline
  \end{tabular}
  \end{center}
  \end{table}

We want to stress in here that the transseries expansion we construct will only be valid for $\real g>0$, simply because the transseries expansion for $r(x)$ has been obtained in this half plane, see equations \eqref{eq:V-U}.
Even if  the Borel-Ecalle resummation of the complete transseries could give us an analytic continuation of the cusp anomalous dimension for $\real g< 0$, 
this will not be the physical cusp anomaly for $\real g< 0$. 
While we know from the weak coupling convergent expansion that $\Gc(g)=\Gc(-g)$,
the analytic continuation of the transseries would break this symmetry!
The correct way to proceed is to obtain the transseries expansion for $r(x)$ for $\real g<0$ and construct from here the new transseries in the left half plane. When we glue together the two Borel-Ecalle resummations of the two transseries for $\real g\gtrless 0$, we will obtain the physical cusp anomaly satisfying $\Gc(g)=\Gc(-g)$, and clearly at the gluing line, the imaginary direction, we will have branch cuts, as expected from weak coupling results.%
\footnote{It is already mentioned in \cite{BES, BBKS} that the $\Gc(g)$ has branch cuts along the imaginary axis starting at $g= \pm \ri /4$.}

Since $r(x_m)$ receives exponentially suppressed corrections in $g \to \infty$, 
the functions $f_0(t)$ and $f_1(t)$ also
receive the non-perturbative corrections of the form $\re^{-2\pi g}$ at strong coupling.
They are thus given by transseries expansions
\be
\ba
f_0(t)&=f_0^{(0)}(t)+\frac{\Lambda^2}{4\pi g}f_0^{(1)}(t)
+\(\frac{\Lambda^2}{4\pi g}\)^{2} f_0^{(2)}(t)+\cO(\Lambda^6), \\
f_1(t)&=f_1^{(0)}(t)+\frac{\Lambda^2}{4\pi g}f_1^{(1)}(t)
+\(\frac{\Lambda^2}{4\pi g}\)^{2} f_1^{(2)}(t)+\cO(\Lambda^6),
\ea\label{eq:f0f1}
\ee
where the non-perturbative scale is given by \eqref{eq:Lambda}.
We want to compute the $1/g$ corrections in  $f_{j}^{(n)}(t)$ ($j=0,1$).
As was computed up to $\cO(1/g^2)$ in \cite{BK2}, the perturbative part takes the following beautiful form
\be
\ba
f_0^{(0)}(4\pi g t)&=\sum_{\ell=0}^\infty \frac{1}{(4\pi g)^\ell} \left[ \gamma_0(t)P_{0,\ell}^{(0)}\( \frac{1}{t} \)
+ \gamma_1(t) Q_{0,\ell}^{(0)} \( \frac{1}{t} \) \right],\\
f_1^{(0)}(4\pi g t)&=\sum_{\ell=0}^\infty \frac{1}{(4\pi g)^\ell} \left[ \gamma_0(t)P_{1,\ell}^{(0)}\( \frac{1}{t} \)
+ \gamma_1(t) Q_{1,\ell}^{(0)} \( \frac{1}{t} \) \right],
\ea
\ee
where
\be
\gamma_0(t)=\frac{\Gamma(\frac{3}{4})\Gamma(1-t)}{\Gamma(\frac{3}{4}-t)},\qquad
\gamma_1(t)=\frac{\Gamma(\frac{1}{4})\Gamma(1+t)}{\Gamma(\frac{1}{4}+t)}.
\ee
A remarkable observation\footnote{We thank Benjamin Basso for telling us this remarkable structure.
Note that a similar property is also found in a different approach in \cite{Volin}.}
is that $P_{j,\ell}^{(0)}(1/t)$ and $Q_{j,\ell}^{(0)}(1/t)$ ($j=0,1$) are \textit{polynomials of $1/t$
with degree $\ell$}.
One can check this observation by computing $f_0^{(0)}(t)$ and $f_1^{(0)}(t)$ order by order, 
following \cite{BK2}.
Here we \textit{assume} this observation in \textit{all orders} in the $1/g$ expansion.
At the leading order: $\ell=0$, we have
\be
P_{0,0}^{(0)}\(\frac{1}{t} \)=-1,\qquad
Q_{0,0}^{(0)}\(\frac{1}{t} \)=P_{1,0}^{(0)}\(\frac{1}{t} \)=Q_{1,0}^{(0)}\(\frac{1}{t} \)=0.
\label{eq:PQ-initial}
\ee
The polynomials $P_{j,\ell}^{(0)}(1/t)$ and $Q_{j,\ell}^{(0)}(1/t)$ ($j=0,1$) are uniquely fixed by the following four
conditions
\begin{itemize}
\item The quantization condition \eqref{eq:QC2}. 
\item The analyticity condition \eqref{eq:f0-t0}.
\item The function $f_1(4\pi g t)$ must be finite at $t=0$ because the cusp anomalous dimension 
\eqref{eq:Gamma-cusp1} is a finite quantity.
\item The polynomials $Q_{0,\ell}^{(0)}(1/t)$, $P_{1,\ell}^{(0)}(1/t)$ and $Q_{1,\ell}^{(0)}(1/t)$ do not have constant terms, i.e.,
$Q_{0,\ell}^{(0)}(0)=P_{1,\ell}^{(0)}(0)=Q_{1,\ell}^{(0)}(0)=0$.
\end{itemize}
Here the last condition is based on the observation in the low order computation.
We do not have a clear reason of this assumption.
We indeed computed the polynomials $P_{j,\ell}^{(0)}(1/t)$ and $Q_{j,\ell}^{(0)}(1/t)$ 
numerically order by order up to $n=180$, starting with \eqref{eq:PQ-initial}.
Once these functions are found, one can immediately compute the strong coupling expansion of the
cusp anomalous dimension by \eqref{eq:Gamma-cusp1}.
We give the explicit results up to $1/g^{10}$ in appendix~\ref{sec:strong}.
Since the solution here is based on the above assumptions, we need to confirm its
validity. 
We have confirmed the strong coupling expansion of $\Gc(g)$ computed in this way is in perfect agreement
with the one computed in another way in \cite{Volin} up to $1/g^{42}$.

In the similar manner, one can fix the non-perturbative corrections $f_0^{(n)}(t)$ and $f_1^{(n)}(t)$ in principle.
We observe that the leading and the next-to-leading corrections take the following form:
\be
\ba
f_0^{(n)}(4\pi g t)&=\sum_{\ell=0}^\infty \frac{1}{(4\pi g)^\ell} \left[ \widehat{\gamma}_0(t)P_{0,\ell}^{(n)}\( \frac{1}{t} \)
+ \widehat{\gamma}_1(t) Q_{0,\ell}^{(n)} \( \frac{1}{t} \) \right],\\
f_1^{(n)}(4\pi g t)&=\sum_{\ell=0}^\infty \frac{1}{(4\pi g)^\ell} \left[ \widehat{\gamma}_0(t)P_{1,\ell}^{(n)}\( \frac{1}{t} \)
+ \widehat{\gamma}_1(t) Q_{1,\ell}^{(n)} \( \frac{1}{t} \) \right], 
\ea
\qquad (n=1,2),
\ee
where
\be
\widehat{\gamma}_0(t)=\gamma_0(t),\qquad
\widehat{\gamma}_1(t)=\frac{\Gamma(\frac{5}{4})\Gamma(1+t)}{\Gamma(\frac{5}{4}+t)}
=\frac{\gamma_1(t)}{1+4t}.
\ee
The functions $P_{j,\ell}^{(n)}(1/t)$ and $Q_{j,\ell}^{(n)}(1/t)$ ($j=0,1;n=1,2$) are again polynomials in $1/t$
with degree $\ell$.
For $\ell=0$ we have
\be
\ba
P_{0,0}^{(1)}\(\frac{1}{t} \)&=\frac{1}{2},\\
P_{0,0}^{(2)}\(\frac{1}{t} \)&=-\frac{1}{4},
\ea
\quad
\ba
Q_{0,0}^{(1)}\(\frac{1}{t} \)&=-\frac{1}{2},\\
Q_{0,0}^{(2)}\(\frac{1}{t} \)&=\frac{1}{4},
\ea
\quad
\ba
P_{1,0}^{(1)}\(\frac{1}{t} \)&=0,\\
P_{1,0}^{(2)}\(\frac{1}{t} \)&=0,
\ea
\quad
\ba
Q_{1,0}^{(1)}\(\frac{1}{t} \)&=1,\\
Q_{1,0}^{(2)}\(\frac{1}{t} \)&=\frac{1}{2},
\ea
\ee
We indeed computed $P_{j,\ell}^{(n)}(1/t)$ and $Q_{j,\ell}^{(n)}(1/t)$ ($n=1,2$) up to $\ell=180$ numerically.%
\footnote{In the non-perturbative sectors, we need to relax the fourth condition above. We require only $P_{1,\ell}^{(n)}(0)=0$ for $n=1,2$.
We confirmed that this condition and the first three conditions above uniquely fix the solution.}
These solutions will be used to analyze the large order behavior of the cusp anomalous dimension in section~\ref{sec:cusp}.

The correction at $\cO(\Lambda^6)$ is much more complicated. As mentioned before, at this order the quantization condition \eqref{eq:QC2}
receives the new contribution from $r(x_1)$.
This additional non-perturbative contribution makes the solution more involved.
In appendix~\ref{sec:3inst}, we work out the leading correction at $\cO(\Lambda^6)$ by following the original method in \cite{BK2}.

\subsection{Numerical evaluation at finite coupling}\label{subsec:numerical}
Here we review how to evaluate the cusp anomalous dimension numerically at finite coupling \cite{BBKS}. 
The basic idea is to reduce the problem from the integral equation to an infinite dimensional linear system.
The two functions $\gamma_\pm(t)$ in \eqref{eq:gamma-pm} admit the following Neumann series:
\be
\ba
\gamma_-(t)=2\sum_{m=1}^\infty(2m-1) \gamma_{2m-1} J_{2m-1}(t),\qquad
\gamma_+(t)=2\sum_{m=1}^\infty(2m) \gamma_{2m} J_{2m}(t),
\ea
\ee
Plugging these expansions into the integral equations \eqref{eq:BES-rewrite0},
one obtains an infinite number of linear equations
\be
\gamma_{n}+\sum_{m=1}^\infty \cK_{nm}\gamma_m=\frac{1}{2}\delta_{n1},\qquad n\geq 1.
\ee
where
\be
\mathcal{K}_{nm}=2m (-1)^{m(n+1)} \int_0^\infty \frac{\rd t}{t} \frac{J_n(2g t) J_m(2gt)}{\re^t-1}.
\ee
Therefore the coefficients $\gamma_n$ is formally given by
\be
\gamma_n=\frac{1}{2}\( \frac{1}{1+\cK} \)_{n1}
\ee
The cusp anomalous dimension is finally given by
\be
\Gc(g)=8g^2 \gamma_1=4g^2 \( \frac{1}{1+\cK} \)_{11}.
\ee
Of course, it is hopeless to compute the inverse matrix $(1+\cK)^{-1}$ analytically at arbitrary coupling,
but in the practical computation it is sufficient to truncate $\cK$ to a finite dimensional matrix.
The convergence is quite rapid as a cut-off of the matrix size grows.
For example, to get the value at $g=1$ with 20-digit precision, it is sufficient to take the $30 \times 30$ truncated matrix.
In table~\ref{tab:cusp-num}, we show the several numerical values of $\Gc(g)/(2g)$ for $1/10\leq g \leq 2$.

\begin{table}[tb]
\caption{The numerical values of the cusp anomalous dimension.}
\label{tab:cusp-num}
\begin{center}
  \begin{tabular}{cc||cc}\hline
$g$ & $\Gc(g)/(2g)$ & $g$ & $\Gc(g)/(2g)$   \\ \hline  
$0.1$ & $0.19385434324817514169$ & $0.2$ & $0.35843733036950590918$ \\
$0.3$ & $0.48610560959171757729$ & $0.4$ & $0.58210341218853482707$ \\
$0.5$ & $0.65393902847754809138$ & $0.6$ & $0.70804410773964657108$ \\
$0.7$ & $0.74930396158229752404$ & $0.8$ & $0.78126610325935190716$ \\
$0.9$ & $0.80645604710807850846$ & $1.0$ & $0.82665913262694175149$ \\
$1.1$ & $0.84313878800192894948$ & $1.2$ & $0.85679498059481188901$ \\
$1.3$ & $0.86827541056713620211$ & $1.4$ & $0.87805219572881873579$ \\
$1.5$ & $0.88647437750685446821$ & $1.6$ & $0.89380385613927175199$ \\
$1.7$ & $0.90024008980819058115$ & $1.8$ & $0.90593719988552639946$ \\
$1.9$ & $0.91101593465281368193$ & $2.0$ & $0.91557213314609428817$ \\
\hline
\end{tabular}
\end{center}
\end{table}

\subsection{The mass gap in the O(6) sigma model}
It is known that the cusp anomalous dimension at strong coupling is closely related to the mass gap in the O(6)
sigma model.
This was first discussed by Alday and Maldacena in the dual string consideration \cite{AM2}.
Then in \cite{BK1,BBBKP, BK2}, the relation was embedded into $\cN=4$ SYM.
In particular, in \cite{BK2}, the mass gap $m_\text{O(6)}$ is exactly related to 
the solution to the BES equation by
\be
m_\text{O(6)}
=\frac{16\sqrt{2}}{\pi^2} \frac{f_1(-\pi g)}{V_0(-\pi g)},\label{eq:massgap}
\ee
where $f_1(t)$ is the same function appearing in the BES solution.
This mass gap scales non perturbatively as $\re^{-\pi g}$ at strong coupling (see \eqref{eq:V0V1}).
Remarkably, the leading non-perturbative correction to the cusp anomalous dimension 
is completely captured by this mass gap \cite{BK2}:
\be
\frac{\Gc(g)}{2g}=\sum_{\ell=0}^\infty \frac{\Gamma_\ell^{(0)}}{(2\pi g)^\ell}
-\frac{\sigma}{8\sqrt{2}g} m_\text{O(6)}^2+\cO(m_\text{O(6)}^4).
\label{eq:cusp-mass}
\ee
The mass gap itself also has the following transseries expansion
\be
\ba
m_\text{O(6)}&=\frac{\sqrt{2}}{\Gamma(\frac{5}{4})}(2\pi g)^{1/4} \re^{-\pi g}
\biggl[ m_\text{O(6)}^{(0)}-\frac{\Lambda^2}{8\pi g}m_\text{O(6)}^{(1)}+\(-\frac{\Lambda^2}{8\pi g} \)^2 m_\text{O(6)}^{(2)}
+\cO(\Lambda^6) \biggr], \\
m_\text{O(6)}^{(n)}&=\sum_{\ell=0}^\infty \frac{m_{\ell}^{(n)}}{(2\pi g)^{\ell}},
\ea
\label{eq:mass-trans}
\ee
where only the very first few exact values of $m_\ell^{(0)}$ and $m_\ell^{(1)}$ are found in \cite{BK2}.
One important consequence of \eqref{eq:cusp-mass} is the following.
The ``perturbative'' coefficients $m_\ell^{(0)}$ are computed by the perturbative BES solution $f_1^{(0)}(t)$ at $t=-\pi g$.
Using the relation \eqref{eq:cusp-mass}, the coefficients $\Gamma_\ell^{(1)}$ are fixed by $m_\ell^{(0)}$.
More explicitly, we have
\be
\sum_{\ell=0}^\infty \frac{\Gamma_\ell^{(1)}}{(2\pi g)^{\ell}}
=\left[ \sum_{\ell=0}^\infty \frac{m_{\ell}^{(0)}}{(2\pi g)^{\ell}} \right]^2.
\label{eq:cusp-mass-1}
\ee
We conclude that all the coefficients $\Gamma_\ell^{(1)}$ can be predicted only from the perturbative
solution to the BES equation.
As we will see in section~\ref{sec:cusp}, we find that a similar relation also holds in the next-to-leading non-perturbative sector
(see \eqref{eq:cusp-mass-2}).

\section{A toy model}\label{sec:toy}

In \cite{BK2}, Basso and Korchemsky constructed an interesting toy model solution of the integral equation (\ref{eq:BES-rewrite}). 
This toy model solution remarkably captures many features that the actual cusp anomalous dimension has.
Thus it is useful to understand $\Gc(g)$ at strong coupling from this toy model, as we will see here in detail.%
\footnote{However, one should keep in mind that there are some significant differences between them. We will comment on them in the next section.}

\subsection{Transseries solution}
The solution is obtained by relaxing the analyticity conditions (\ref{eq:QC}).
The toy model solution does not have any poles, but has only the zero at $t=- \pi \ri \,g$.
This analyticity condition can be easily imposed directly in equation (\ref{eq:Gammahat}) and one finds
\be
\ba
 f_0^\text{(toy)}(t)&=-1,\\
 f_1^\text{(toy)}(t)&=-c^\text{(toy)}(g),\qquad c^\text{(toy)}(g)=-\frac{V_0(-\pi g)}{V_1(-\pi g)}.
\ea
\ee
Surprisingly, both  $f_0^\text{(toy)}(t)$ and $ f_1^\text{(toy)}(t)$ do not depend on $t$.
The exact solution to the BES equation thus takes the form
\be
\ba
\widehat{\Gamma}^\text{(toy)}(\ri t)&=-V_0(t)-c^\text{(toy)}(g) V_1(t),\\
\ea
\ee
where $V_0(t)$ and $V_1(t)$ are the same functions as before, and 
can be expressed in terms of Whittaker functions (see \eqref{eq:Vn-M}) .

From the above expression we can obtain the toy model cusp anomalous dimension
\be
\ba
\Gc^\text{(toy)}(g)&=-2g \widehat{\Gamma}^\text{(toy)}(0)=2g \left[1-2\frac{V_0(-\pi g)}{V_1(-\pi g)} \right] \\
&=2g \left[ 1-\frac{1}{\sqrt{2\pi g}} \frac{M_{1/4,1/2}(2\pi g)}{M_{-1/4,0}(2\pi g)} \right].
\ea
\ee
where we used the relation \eqref{eq:Vn-M} to the Whittaker function of the first kind.
Similarly to the true cusp anomalous dimension, the small coupling expansion in the toy model also has a finite radius of convergence
$\vert g \vert <  0. 7966\dots$, while the strong coupling expansion is only asymptotic as we will shortly prove.

Using the result (\ref{eq:V0V1}) derived in the appendix we can rewrite the toy model cusp anomaly
\be
\frac{\Gc^\text{(toy)}(g)}{2g}=1-\alpha \frac{\cS_{+}F(\frac{1}{4},\frac{5}{4}|\alpha)+\Lambda_+^2 \cS_0F(-\frac{1}{4},\frac{3}{4}|{-\alpha})}
{\cS_{+} F(\frac{1}{4},\frac{1}{4}|\alpha)+\frac{1}{4}\Lambda_+^2 \alpha \, \cS_0 F(\frac{3}{4},\frac{3}{4}|{-\alpha})},\qquad
\alpha=\frac{1}{2\pi g},
\label{eq:Gamma-toy}
\ee
where $F(a,b|z)$ is defined by the asymptotic series \eqref{eq:F-def}, and 
$\cS_{\pm}=\cS_{0^\pm}$ are the lateral Borel resummations along the $\theta=0$ direction.
The Borel resummation of $F(a,b|z)$ is defined by \eqref{eq:Borel-F}.
Similarly, it is also possible to write it as
\be
\frac{\Gc^\text{(toy)}(g)}{2g}=1-\alpha \frac{\cS_- F(\frac{1}{4},\frac{5}{4}|\alpha)+\Lambda_-^2 \cS_0 F(-\frac{1}{4},\frac{3}{4}|{-\alpha})}
{\cS_- F(\frac{1}{4},\frac{1}{4}|\alpha)+\frac{1}{4}\Lambda_-^2 \alpha\, \cS_0 F(\frac{3}{4},\frac{3}{4}|{-\alpha})},
\label{eq:Gamma-toy2}
\ee
where $\Lambda_\pm$ are the non-perturbative scales
\be
\Lambda_\pm^2
=\sigma_\pm (2\pi g)^{1/2} \re^{-2\pi g}, \qquad
\sigma_\pm = \re^{\mp 3\pi \ri/4} \frac{\Gamma(\frac{3}{4})}{\Gamma(\frac{5}{4})}.
\label{eq:Lambda-pm}
\ee
Note that the expressions \eqref{eq:Gamma-toy} and \eqref{eq:Gamma-toy2} are equivalent, 
and both Borel resummation procedures $\cS_\pm$ leads to the same
result unambiguously thanks to the non-perturbative parts $\Lambda_\pm^2$.

We can now expand (\ref{eq:Gamma-toy}) or (\ref{eq:Gamma-toy2}) at strong coupling (i.e. for $\alpha$ small with $\mbox{Re}\,\alpha>0$) and obtain the transseries expansion for the toy model cusp anomaly
\begin{equation}
\frac{\Gc^\text{(toy)}(g)}{2g}= 
\left\lbrace
\begin{matrix}C_0(\alpha)-\alpha \Lambda_+^2 C_2(\alpha)+\frac{1}{4} \alpha^2 \Lambda_+^4 C_4(\alpha)+\cO(\Lambda_+^6)\,,\qquad 0<\mbox{arg} \,\alpha<\frac{\pi}{2}\,,\\
 \,C_0(\alpha)-\alpha \Lambda_-^2 C_2(\alpha)+\frac{1}{4} \alpha^2 \Lambda_-^4 C_4(\alpha)+\cO(\Lambda_-^6)\,,\qquad-\frac{\pi}{2}<\mbox{arg} \,\alpha<0\,,\end{matrix}\right.
 \label{eq:ToyTS}
\end{equation}
where the perturbative expansion $C_0(\alpha)$ is 
\begin{equation}
C_0(\alpha)=1-\alpha \frac{F(\frac{1}{4},\frac{5}{4}|\alpha)}{F(\frac{1}{4},\frac{1}{4}|\alpha)}\,,\label{eq:C0}
\end{equation}
while each function $C_{2n}(\alpha)$ is given by
\be
\ba
C_2(\alpha)=\frac{1}{F(\frac{1}{4},\frac{1}{4}|\alpha)^2},\quad
C_4(\alpha)=\frac{F(\frac{3}{4},\frac{3}{4}|{-\alpha})}{F(\frac{1}{4},\frac{1}{4}|\alpha)^3},\quad
C_{2n}(\alpha)=\frac{\left(F(\frac{3}{4},\frac{3}{4}|{-\alpha})\right)^{n-1}}{\left(F(\frac{1}{4},\frac{1}{4}|\alpha)\right)^{n+1}}.
\ea
\label{eq:C2n}
\ee
To obtain these results, we used the Wronskian like relation \eqref{eq:Wronskian}.
Given the series expansion (\ref{eq:F-def}), it is clear that each $C_{2n}(\alpha)$, once expanded for small $\alpha$, is given by an asymptotic power series, non-Borel summable for $\alpha\in\mathbb{R}^+$.
Note however that $F(\frac{3}{4},\frac{3}{4}|{-\alpha})$ is instead Borel summable for real and positive $\alpha$ while it becomes non-Borel summable\footnote{This means that, while for $C_0$ and $C_2$ the only Stokes line is $\arg \alpha = 0$, for higher $C_{2n}$ also $\arg \alpha =\pi $ will be a Stokes line.} for $\alpha\in\mathbb{R}^-$.

Following the results derived in Section \ref{sec:Borel}, the Borel-Ecalle resummation of the transseries (\ref{eq:ToyTS}) is given by
\begin{equation}
\frac{\Gc^\text{(toy)}(g)}{2g}= 
\left\lbrace
\begin{matrix}\mathcal{S}_+ C_0(\alpha)-\alpha \Lambda_+^2 \mathcal{S}_+C_2(\alpha)+\frac{1}{4} \alpha^2 \Lambda_+^4 \mathcal{S}_+ C_4(\alpha)+\cdots\,,\,\, 0<\mbox{arg} \,\alpha<\frac{\pi}{2}\,,\\
\, \mathcal{S}_- C_0(\alpha)-\alpha \Lambda_-^2 \mathcal{S}_- C_2(\alpha)+\frac{1}{4} \alpha^2 \Lambda_-^4 \mathcal{S}_- C_4(\alpha)+\cdots\,,\,\,-\frac{\pi}{2}<\mbox{arg}\, \alpha<0\,.\end{matrix}\right.\,
 \label{eq:ToyTSResummed}
\end{equation}
Despite the direction $\arg \alpha = 0$ being a Stokes line for both $C_0$ and all the $C_{2n}$, the resummation of the transseries expansion (\ref{eq:ToyTSResummed}) is perfectly good and analytic in the entire right half-plane $-\pi/2 < \mbox{arg}\,\alpha<\pi/2$. As we will show, the ambiguity in the resummation prescription $(\mathcal{S}_+ -\mathcal{S}_-)C_0\neq 0 $, and similarly  $(\mathcal{S}_+ -\mathcal{S}_-)C_{2n}\neq 0 $, is exactly balanced by the jump in the transseries parameter $\sigma_+ \sim \re^{-3\pi \ri/4} \to \sigma_- \sim \re^{+3\pi \ri/4}$.

Despite the fact that the Borel-Ecalle resummation of the transseries (\ref{eq:ToyTSResummed}) defines an analytic continuation for $\Gc^\text{(toy)}(g)$ in the entire complex plane, minus the negative real axis, nonetheless this is not the correct analytic continuation for $\mbox{Re}\, g < 0$. Similarly to our previous discussion in Section \ref{sec:BES}, this transseries ansatz is only a solution to the toy model quantization conditions in the right half plane $\mbox{Re}\, g > 0$, since we used the transseries form (\ref{eq:V0V1}) for $V_0$ and $V_1$ only valid in this half plane. We should first obtain the correct transseries in the left half plane $\mbox{Re}\, g <0$, and then perform its Borel-Ecalle resummation to obtain the toy model cusp anomaly.
In particular the correct analytic continuation for $\Gc^\text{(toy)}(g)$ will not have a branch cut on the negative real axis similar to the true $\Gc$, as one can easily see from the weak coupling analysis.

Naively, to compute the discontinuity $(\mathcal{S}_+ -\mathcal{S}_-)C_0$,  one expands (\ref{eq:C0}) for small $\alpha$, 
computes its Borel transform, and then studies the analyticity properties of $\mathcal{B} [C_0]$.
However, it is much better to use the fact that $\mathcal{S}_\pm$ are good resummation prescriptions, 
i.e.,
\be
\mathcal{S}_+ C_0(\alpha) = \mathcal{S}_+\left(1-\alpha \frac{F(\frac{1}{4},\frac{5}{4}|\alpha)}{F(\frac{1}{4},\frac{1}{4}|\alpha)}\right)=1-\alpha \frac{\mathcal{S}_+ F(\frac{1}{4},\frac{5}{4}|\alpha)}{\mathcal{S}_+ F(\frac{1}{4},\frac{1}{4}|\alpha)}\,.
\ee
One can indeed check this equality by performing the lateral Borel(-Pad\'e) resummation on both hand sides.

At this point we can make use of the trivial identity $\mathcal{S}_+= \mathcal{S}_-+(\mathcal{S}_+-\mathcal{S}_-)$, together with the known discontinuity
 (\ref{eq:disc}) for $F(a,b|\alpha)$, to obtain
\be
\mathcal{S}_+ C_0(\alpha) = \mathcal{S}_- C_0(\alpha)+4\sum_{n=1}^\infty \left(-\frac{\ri \alpha\,\Lambda_0^2}{2\sqrt{2}}\right)^n \mathcal{S}_-C_{2n}(\alpha)\,,\label{eq:DiscC0}
\ee
with $\Lambda_0^2 =  \re^{\pm3\pi \ri/4}\Lambda_{\pm}^2$.
The discontinuity of the resummation of the perturbative expansion is precisely of the form (\ref{eq:Disc}), discussed in Section \ref{sec:Borel}.
Similar results can be derived for all higher perturbative corrections to the non-perturbative terms $\Lambda_{\pm}^{2n}$, all schematically of the form
\be
\mathcal{S}_+ C_{2n}(\alpha) = \mathcal{S}_- C_{2n}(\alpha) -(n+1) \frac{\ri \alpha \Lambda_0^2}{2\sqrt{2}}\, \mathcal{S}_-C_{2(n+1)} (\alpha)\,+O(\Lambda_0^4)\label{eq:DiscCn}
\ee
We can now understand why the Borel-Ecalle resummation (\ref{eq:ToyTSResummed}), despite looking discontinuous for $\arg \alpha = 0$, defines indeed an analytic function in the right half-plane: the change in the transseries parameter in front of the $C_2$ term is taking care of the non-perturbative contribution of order $\Lambda_0^2$ in the discontinuity (\ref{eq:DiscC0}) of $C_0$. The jump of the $C_4$ coefficients balances the sum of the non-perturbative $\Lambda_0^4$ term in the discontinuity of $C_0$ together with the non-perturbative contribution of order $\Lambda_0^2$ coming from the discontinuity of $C_2$ (\ref{eq:DiscCn}), and so on.

Furthermore, from the discontinuity equation (\ref{eq:DiscC0}) for $C_0(\alpha)$, we can use the result (\ref{eq:HigherOrderP}) for the large order coefficients%
\footnote{Unlike the case in the previous section, there is a fractional exponent factor $\alpha^{-1/2}$ in $\Lambda_0^2$.
This factor shifts the arguments of the Gamma function, but the essential argument does not change at all.}
 of the perturbative expansion to derive
\be
\ba
C_{0,\,(n)}\sim& \frac{2\, A}{\pi\ri\,(1)^{n-1/2}} \Gamma\left(n-\frac{1}{2}\right) \left(1+ \frac{C_{2,\,(1)}\times 1}{n-\frac{3}{2}}+\frac{C_{2,\,(2)}\times 1^2}{(n-\frac{3}{2})(n-\frac{5}{2})}+ O(n^{-3}) \right)+\\
& \frac{2\,A^2}{\pi\ri\,(2)^{n-1}} \Gamma\left(n-1\right) \left(1+ \frac{C_{4,\,(1)}\times 2}{n-2}+\frac{C_{4,\,(2)}\times 2^2}{(n-2)(n-3)}+ O(n^{-3}) \right)+\\
&\label{eq:LargeOrdC0}+ \frac{2\,A^3}{\pi\ri\,(3)^{n-3/2}} \Gamma\left(n-\frac{3}{2}\right) \left(1+ \frac{C_{6,\,(1)}\times 3}{n-\frac{5}{2}}+O(n^{-2})\right)+\cdots\,,
\ea
\ee
where $C_{0,\,{(n)}},C_{2,\,{(n)}},...$ are the $n$-th order perturbative coefficients of the small $\alpha$ expansion of $C_0,C_2,...$, while the Stokes constant $A$ is simply given by
\be
 A = \frac{-\ri \,\Gamma\left(\frac{3}{4}\right) }{ 2 \sqrt{2} \,\Gamma\left(\frac{5}{4}\right)}
 =\frac{\ri\,\mbox{Im}\, \sigma_+}{2}=\frac{-\ri\,\mbox{Im} \,\sigma_-}{2}\,\,,\label{eq:toyStokes}
 \ee
as one could have also deduce from the jump in the transseries parameter (see \cite{Dorigoni:2014hea}).

As mentioned in Section \ref{sec:Borel}, the large order behaviour (\ref{eq:LargeOrdC0}) of the perturbative coefficients $C_{0,\,(n)}$ can be used to test the validity of our transseries expansion.
From equation (\ref{eq:C0}), we can easily generate 
an arbitrary number of perturbative coefficients $C_{0,\,(n)}$, and check the leading ``one instanton''%
\footnote{In this paper, we sometimes use the term ``instanton'' to count the non-perturbative order $\Lambda^{2}$ for notational simplicity.
Keep in mind that this does not mean that such corrections are caused by instantons in the O(6) sigma model.
There is no instanton configuration in this model, and the Borel singularities correspond to IR renormalons.} 
behaviour $C_{0,\,(n)} \sim A \Gamma(n-1/2)$. 
We can proceed by considering the first $1/n$ correction to the leading $\Gamma(n-1/2)$ term, this will give us the first perturbative coefficient, $C_{2,\,(1)}$, of the one instanton sector, that can be checked against the analytic value obtained from the expansion of (\ref{eq:C2n}). 
Higher and higher terms $C_{2,\,(2)},C_{2,\,(3)},...$ can be obtained by simply looking at higher $n^{-2},n^{-3},...$ corrections.

We already know that the $C_2$ expansion is also asymptotic, so the coefficients $C_{2,\,(n)}$ will diverge as well. For this reason we can perform a Borel-Pade resummation of the leading one instanton sector, $1\mbox{Inst}\sim A \,\Gamma(n-1/2)( 1 + C_{2,\,(1)}/(n-1/2)+...)$, and subtract it from the perturbative coefficient $C_{0,\,(n)}$. In this way we can isolate the subleading large order behavior (\ref{eq:LargeOrdC0}) which is $C_{0,\,(n)} - 1\mbox{Inst} \sim A^2\,  \Gamma(n-1)/ 2^n$. From this new set of coefficients we can read the perturbative coefficients $C_{4,\,(n)}$ of the two-instantons sector, repeat the Borel-Pad\'e resummation and isolate the three-instantons sector and so on.\footnote{We thank Marcel Vonk for useful discussions on this problem.}

We checked numerically, up to the four-instantons sector, that the large orders of the perturbative series expansion do indeed contain all the non-perturbative information, consistently with our one parameter transseries expansion (\ref{eq:ToyTS}).

\subsection{The non-perturbative mass scale}\label{sec:NPmassToy}

In Section \ref{sec:BES}, we briefly discussed the relation between the non-perturbative scale of the cusp anomalous dimension and the mass gap of O(6) sigma model \cite{AM2}.
A similar analysis \cite{BK2} can be carried out for the toy model as well. 
In the toy model the non-perturbative mass is exactly given by
\be
m_\text{toy}=\frac{16\sqrt{2}}{\pi^2} \frac{1}{V_1(-\pi g)}.
\ee
Using the results derived in the Appendix (\ref{eq:V0V1}), we can rewrite the mass gap as
\be
m_\text{toy}=\frac{4}{\pi \Gamma(\frac{5}{4})}\alpha^{-1/4}\re^{-\frac{1}{2\alpha}}
\frac{1}{\cS_{\pm}F(\frac{1}{4},\frac{1}{4}|\alpha)+\frac{1}{4}\alpha \Lambda_\pm^2 \cS_0 F(\frac{3}{4},\frac{3}{4}|{-\alpha})}.
\label{eq:mass-toy}
\ee
Note that in the toy model the mass gap takes a particularly simple transseries form, we just need to expand equation (\ref{eq:mass-toy}) around 
$\Lambda_{\pm}= 0 $, and we obtain a one parameter transseries. The only reason why we get infinitely many non-perturbative corrections in $m_\text{toy}$ is that we expand $1/V_1(-
\pi g)$, had we considered $1/m_\text{toy} \sim  V_1(-\pi g)$ we would have obtained a very simple two-terms transseries (see \cite{Dorigoni:2014hea}).

As in \eqref{eq:ToyTS}, we expand the mass gap as
\begin{equation}
m_\text{toy}= \frac{4}{\pi \Gamma(\frac{5}{4})}\alpha^{-1/4}\re^{-\frac{1}{2\alpha}} \biggl[
m_{\text{toy},0}(\alpha)-\frac{\alpha \Lambda_+^2}{4} m_{\text{toy},1}(\alpha)
+\frac{ \alpha^2 \Lambda_+^4 }{16}m_{\text{toy},2}(\alpha)+\cO(\Lambda_+^6)
\biggr]
\label{eq:mass-toyTS}
\end{equation}
for $0<\mbox{arg} \,\alpha<\frac{\pi}{2}$. For $0<\mbox{arg} \,\alpha<\frac{\pi}{2}$, one replaces $\Lambda_+ \to \Lambda_-$.
The coefficient $m_{\text{toy},n}$ is given by
\be
m_{\text{toy},n}(\alpha)=\frac{[F(\frac{3}{4},\frac{3}{4}| {-\alpha})]^n}{[F(\frac{1}{4},\frac{1}{4}| \alpha)]^{n+1}}, \quad (n \geq 0).
\ee
It is easy to see that there are non-trivial relations between $C_{2n}(\alpha)$ and $m_{\text{toy},n}$,
\be
\ba
\frac{m_{\text{toy},n}(\alpha)}{m_{\text{toy},n-1}(\alpha)}&=\frac{m_{\text{toy},1}(\alpha)}{m_{\text{toy},0}(\alpha)}
=\frac{F(\frac{3}{4},\frac{3}{4}| {-\alpha})}{F(\frac{1}{4},\frac{1}{4}| \alpha)}, \\
C_{2n}(\alpha)&=m_{\text{toy},0}(\alpha) m_{\text{toy},n-1}(\alpha).
\ea
\ee
The latter relation with $n=1$ is the same as the one in \eqref{eq:cusp-mass-1}.
In the toy model, the similar relations to \eqref{eq:cusp-mass-1} hold for all the non-perturbative sectors.
In the next section, we also confirm that the similar relation hold for $n=2$ in the true cusp anomalous dimension.

\section{The cusp anomalous dimension at strong coupling}\label{sec:cusp}

In this section, we study the strong coupling expansion of the physical cusp anomalous dimension.
We first see singularities of the Borel transforms.
Then we analyze the large order behavior of the perturbative expansion in detail.
The result shows that the information on the non-perturbative sectors are encoded in the perturbative
sector.
We next perform the lateral Borel resummation, and compare it with numerical values of $\Gc(g)$
computed directly from the BES equation.

\subsection{Borel singularities}
Let us first see the singularity structure of the Borel transform.
The cusp anomalous dimension has the transseries expansion \eqref{eq:cusp-trans}.
Using the method reviewed in section~\ref{sec:BES}, we have computed the numerical values of
$\Gamma_\ell^{(n)}$ ($n=0,1,2$) up to $\ell=180$ with 200-digit precision.
A very first few values are in perfect agreement with the exact ones in \cite{BKK, BK2}.
Some higher order corrections are presented in appendix~\ref{sec:strong}.

Since $\Gamma_\ell^{(n)}$ grows as $\Gamma(\ell-1/2)$ in $\ell \to \infty$, 
it is natural to consider the following a bit modified Borel transform
\be
\widetilde{\cB}[\Gc^{(n)}](\zeta):=\sum_{\ell=1}^\infty \frac{\Gamma_\ell^{(n)}}{\Gamma(\ell-\frac{1}{2})} \zeta^\ell 
\label{eq:Borel-Gamma}
\ee
Since we have only the finite number of coefficients $\Gamma_\ell^{(n)}$ up to $\ell=180$,
we need an approximation of $\widetilde{\cB}[ \Gc^{(n)} ](\zeta)$.
A natural way is to replace it by its Pad\'e approximant.
In the following, we use the diagonal Pad\'e approximant of $\widetilde{\cB}[ \Gc^{(n)} ](\zeta)$ with order $90$.

\begin{figure}[tb]
\begin{center}
\begin{tabular}{cc}
\resizebox{65mm}{!}{\includegraphics{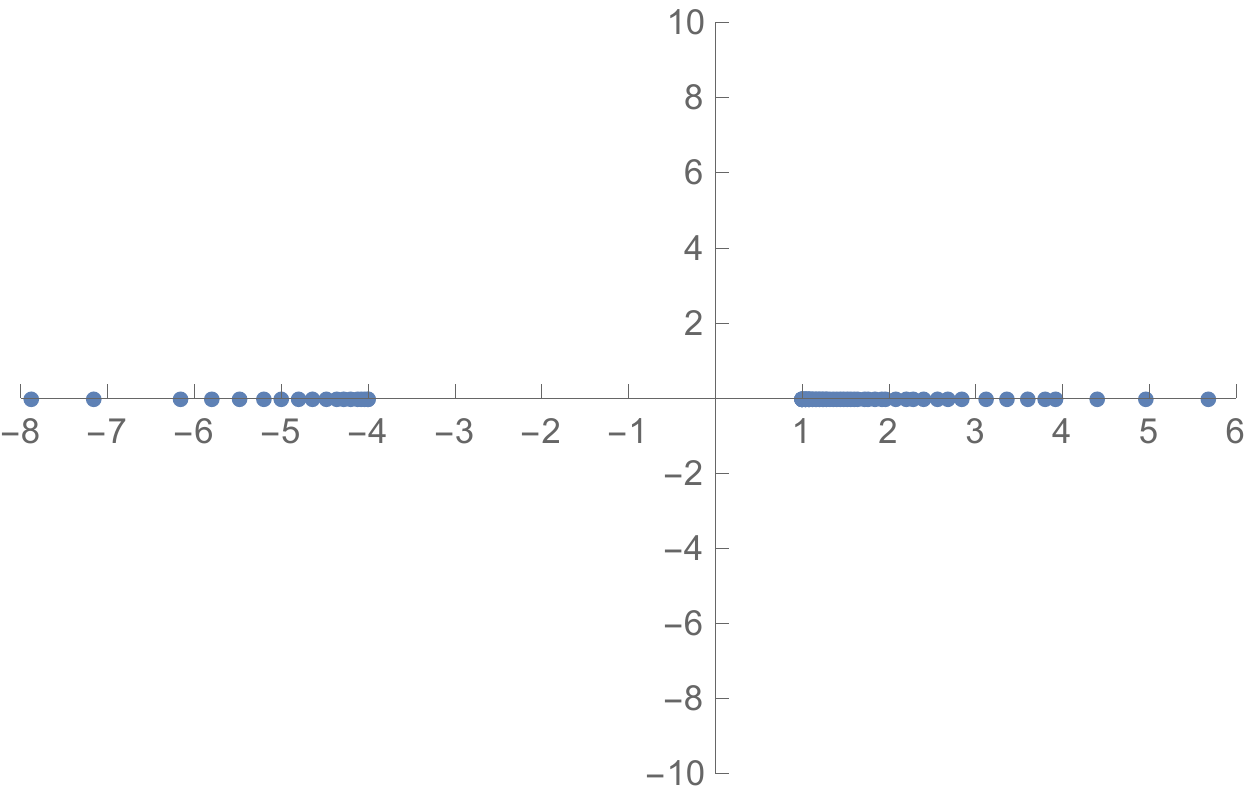}}
\hspace{2mm}
&
\resizebox{65mm}{!}{\includegraphics{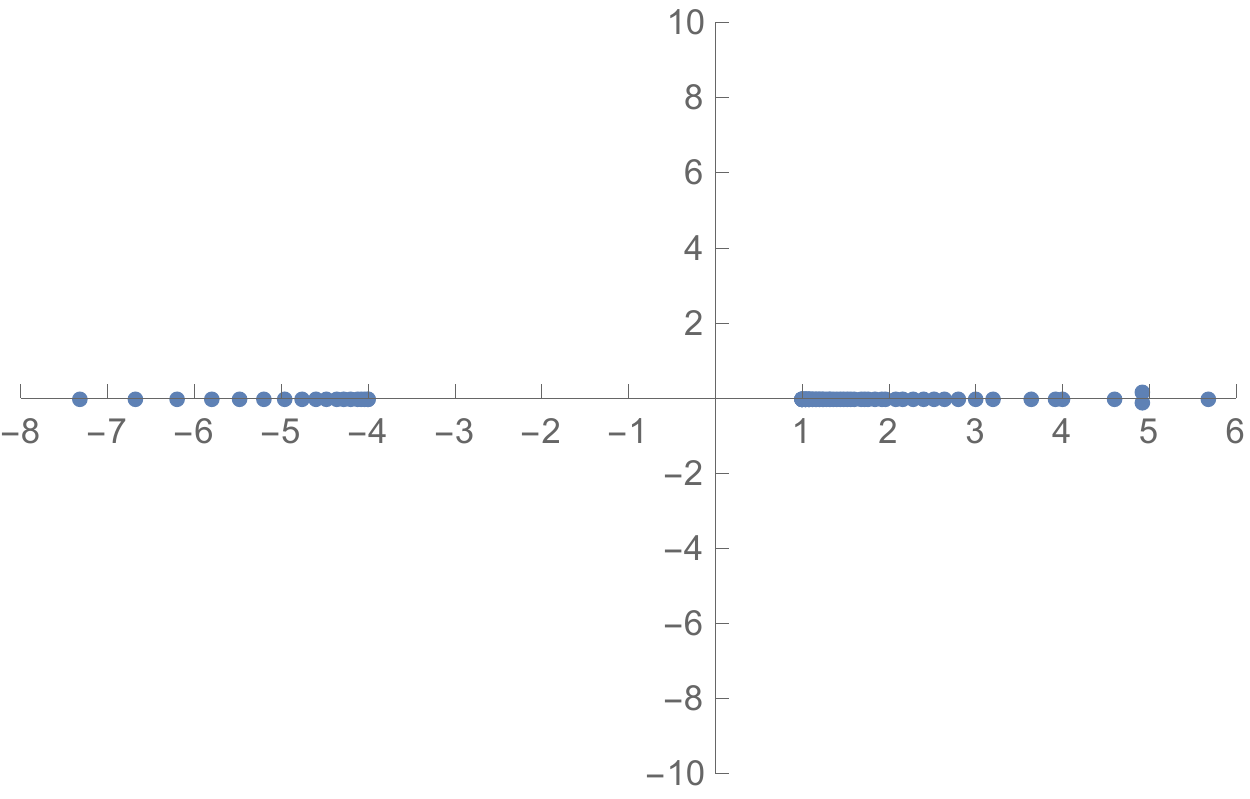}}\\
{\small $n=0$} & {\small $n=1$}
\vspace{5mm}
\end{tabular}
\begin{tabular}{c}
\resizebox{65mm}{!}{\includegraphics{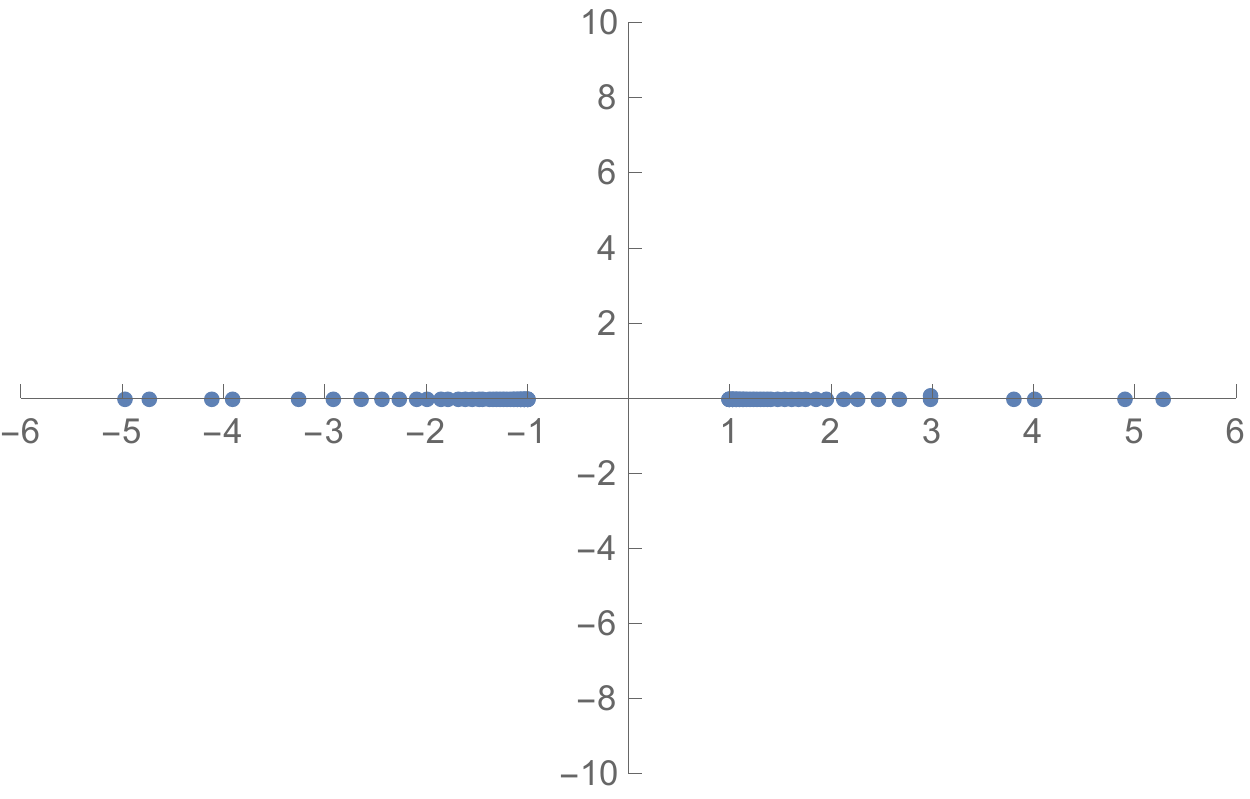}} \\
{\small $n=2$}
\vspace{-3mm}
\end{tabular}
\end{center}
  \caption{We show the singularities of the Pad\'e approximant for the Borel transform $\widetilde{\cB}[ \Gc^{(n)} ](\zeta)$
($n=0,1,2$) in the complex Borel plane. 
It is obvious to see that the Borel transforms for $n=0,1$ have the singularities at $\zeta=1, -4$,
while the Borel transform for $n=2$ has the singularities at $\zeta=\pm 1$. 
}
\label{fig:Borel-sing}
\end{figure}

In figure~\ref{fig:Borel-sing}, we show the Borel singularities of the Pad\'e approximant for $\widetilde{\cB}[ \Gc^{(n)} ](\zeta)$
($n=0,1,2$). The figure clearly shows that the Borel transforms $\widetilde{\cB}[ \Gc^{(n)} ](\zeta)$ for $n=0,1$ have 
the singularities at $\zeta=1,-4$, while $\widetilde{\cB}[ \Gc^{(2)} ](\zeta)$ has the singularities at $\zeta=\pm 1$.
Note that in a Pad\'e approximant, a condensation of poles indicates a branch cut of the original function.
Therefore it is very likely that $\widetilde{\cB}[ \Gc^{(n)} ](\zeta)$ for $n=0,1$ has two branch cuts 
$(-\infty,-4)$ and $(1,\infty)$ while $\widetilde{\cB}[ \Gc^{(2)} ](\zeta)$ has branch cuts $(-\infty, -1)$ and $(1,\infty)$.

These singularity structures are very important to understand the form of the full transseries,
as emphasized in section~\ref{sec:Borel}.
The large order behavior of $\Gamma_\ell^{(n)}$ also heavily depends on the Borel singularities. 

Note that from figure \ref{fig:Borel-sing} it is manifest that also the direction $\theta = \pi$ in the complex Borel plane is a Stokes direction. Following the discussion of Section \ref{sec:Borel} it is clear that from the large order behavior of the perturbative coefficients we will see the presence of these discontinuities. However by the point of view of the cusp anomalous dimension, this additional Stokes line is not relevant. In fact as we have already reviewed in Section \ref{sec:BES}, the transseries that we are studying, is only valid in the right half plane.
It is not necessary to add a transseries parameter to take into account the new singularity at $\zeta=-4$, and the Stokes line $\theta = \pi$ because it is outside of the regime in which the Borel-Ecalle resummation of our transseries reproduces the physical cusp anomaly. Of course the large order behaviour of the perturbative coefficients will know nonetheless of this additional singularity in the negative axis of the Borel plane.

\subsection{Large order behavior}\label{subsec:LOB}
We will study now the large order behavior of the perturbative coefficients.
First of all, as was observed in \cite{BKK}, the perturbative coefficients $\Gamma_\ell^{(0)}$
grow factorially as $\Gamma(\ell-1/2)$ in the large $\ell$ limit.
As in section~\ref{sec:Borel}, the large order behavior of $\Gamma_\ell^{(0)}$
has more information.
It contains the non-perturbative information.
In fact, following the argument in section~\ref{sec:Borel},
it is expected that $\Gamma_\ell^{(0)}$ 
behaves in the large $\ell$ limit as%
\footnote{As shown in figure~\ref{fig:Borel-sing}, the Borel transform $\widetilde{\cB}[\Gc^{(0)}](\zeta)$ 
has the singularity at $\zeta=-4$.
This singularity implies a contribution of order $(-4)^{-\ell}$ to the large order behavior.
Therefore this contribution is important to see $\Gamma_\ell^{(4)}$ from the large order behavior.}
\be
\ba
\Gamma_{\ell}^{(0)}&=\mathcal{A}_1 \Gamma\(\ell-\frac{1}{2}\)
\biggl[1+\frac{\Gamma_1^{(1)}}{\ell-\frac{3}{2}}
+\frac{\Gamma_2^{(1)}}{(\ell-\frac{3}{2})(\ell-\frac{5}{2})}
+\cdots \biggr] \\
&\quad+\mathcal{A}_2 \frac{\Gamma(\ell-1)}{2^{\ell-1}} \biggl[ 1+\frac{2\Gamma_1^{(2)}}{\ell-2}+\frac{2^2 \Gamma_2^{(2)}}{(\ell-2)(\ell-3)}+\cdots \biggr] \\
&\quad+\mathcal{A}_3 \frac{\Gamma(\ell-\frac{3}{2})}{3^{\ell-\frac{3}{2}}}
\biggl[ 1+\cdots 
\biggr]
+\cdots, 
\ea
\qquad\ell \to \infty.
\label{eq:LOB-pert}
\ee
where the first two coefficients $\mathcal{A}_1$ and $\mathcal{A}_2$ are 
exactly the same as the toy model ones in \eqref{eq:LargeOrdC0}-(\ref{eq:toyStokes}):
\be\ba
\mathcal{A}_1&=\frac{2A}{\pi \ri}
=-0.3042971194498708318670259057666\dots. \\
\mathcal{A}_2&=\frac{2A^2}{\pi \ri}=0.14545061420433549651130121616\dots \ri.
\label{eq:A0}
\ea\ee
As we will see later, the constant $\mathcal{A}_3$ is more subtle.
Below we want to confirm the relation \eqref{eq:LOB-pert} by using the numerical values of $\Gamma_\ell^{(0)}$, $\Gamma_\ell^{(1)}$ and
$\Gamma_\ell^{(2)}$ computed from the BES solution.

We first estimate $\mathcal{A}_1$.
Let us consider a sequence
\be
a_\ell^{(1)}=\frac{\Gamma_\ell^{(0)}}{\Gamma(\ell-\frac{1}{2})}.
\ee
This sequence must converge to $\mathcal{A}_1$ in $\ell \to \infty$.
The naive evaluation of $a_\ell^{(1)}$ at $\ell=180$ shows
\be
a_{180}^{(1)} \approx -0.304049.
\ee
This is indeed close to the exact value in \eqref{eq:A0},
but the agreement is not so good because of the sub-leading contribution.
In fact, the sub-leading contribution is of order $1/\ell \sim 5\times 10^{-3}$.
To improve the precision, we need to remove such corrections.
This can be done by using the Richardson extrapolation.
Let us consider the $n$-th Richardson transform \cite{Marino:2007te} of a sequence $f_\ell$,
\be
\cR_n [f_\ell]:=\sum_{k=0}^n \frac{ (-1)^{k+n}(\ell+k)^n}{k! (n-k)!}f_{\ell+k}.
\ee
Assuming \eqref{eq:LOB-pert}, the sequence $a_\ell^{(1)}$ behaves as
\be
a_\ell^{(1)}=\mathcal{A}_1 [1+\cO(\ell^{-1}) ], \qquad \ell \to \infty.
\ee
The Richardson transform of $a_\ell^{(1)}$ then behaves as
\be
\cR_n [a_\ell^{(1)}]=\mathcal{A}_1 [1+\cO(\ell^{-n-1}) ], \qquad \ell \to \infty.
\ee
Therefore the convergence speed is improved.
Note that to compute the $n$-th Richardson transform of $a_\ell^{(1)}$ we need the higher coefficients 
$a_{\ell+1}^{(1)},\dots, a_{\ell+n}^{(1)}$.
Thus if we have $a_\ell^{(1)}$ up to certain finite $\ell=\ell_\text{max}$,
we can perform the $n$-th Richardson transform only up to $\ell'=\ell_\text{max}-n$.
It is a trade-off how we should choose $n$ and $\ell'$.

\begin{figure}[tb]
\begin{center}
\begin{tabular}{cc}
\resizebox{70mm}{!}{\includegraphics{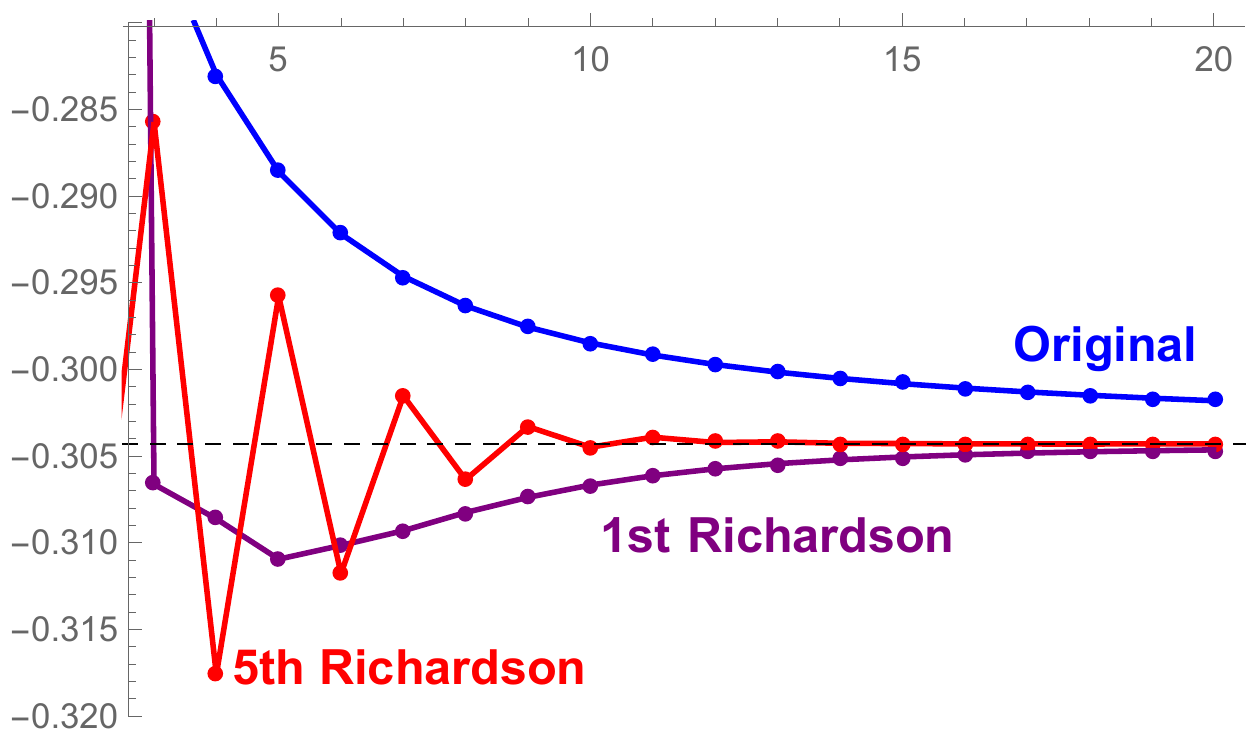}}
\hspace{3mm}
&
\resizebox{70mm}{!}{\includegraphics{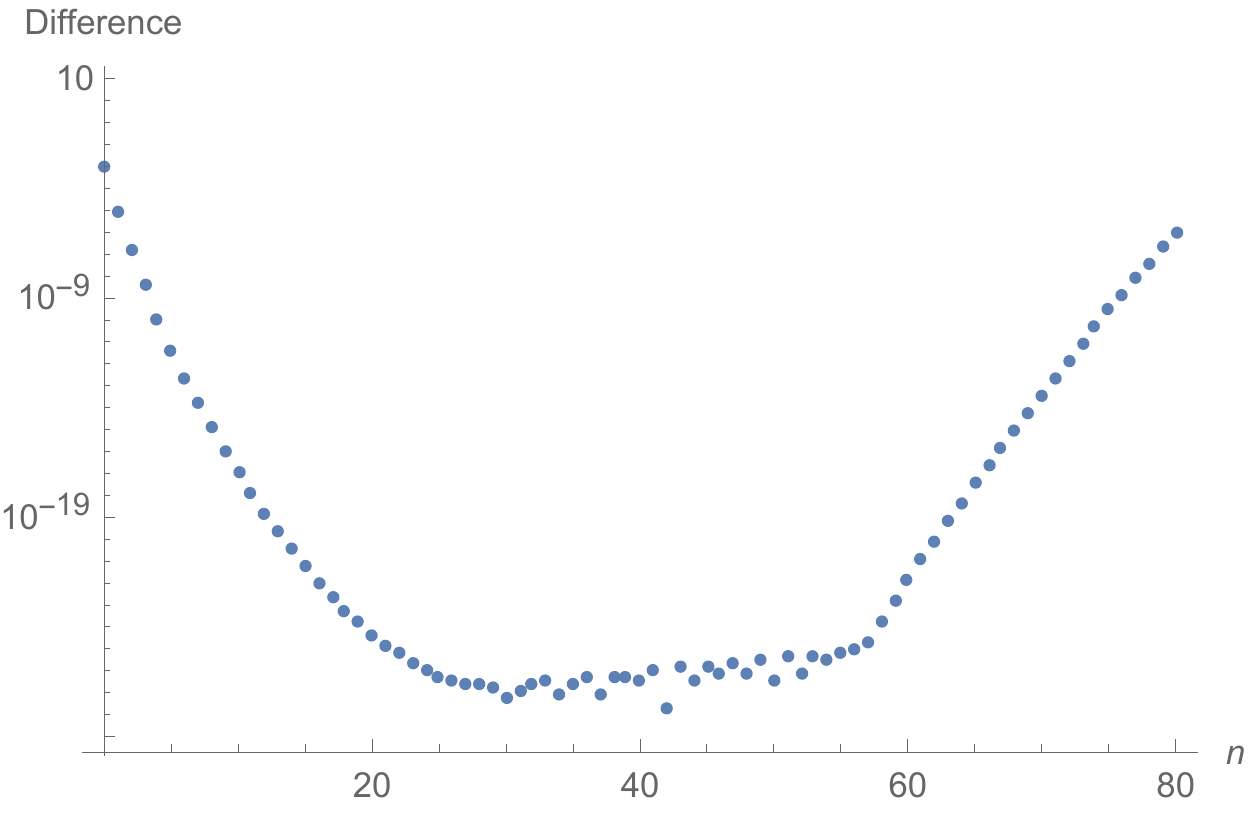}}
\end{tabular}
\end{center}
  \caption{(Left) We show the convergence behavior of the original sequence $a_\ell^{(1)}$, 
its first and fifth Richardson transforms by the blue, purple and red curves, respectively.
The black dashed line is the expected convergence value $\mathcal{A}_1$.
The Richardson extrapolation accelerates the convergence.
(Right) The difference $|1 - \cR_n [a_{180-n}^{(1)}]/\mathcal{A}_1|$ is shown as a function of $n$.
In this case, the Richardson extrapolation is especially good for the range $20 \lesssim n \lesssim 60$.
}
\label{fig:Richardson}
\end{figure}

In the left of figure~\ref{fig:Richardson}, we sketch the convergence behavior of the Richardson extrapolation 
for the sequence $a_\ell^{(1)}$.
We plot the original $a_\ell^{(1)}$, its first and fifth Richardson transforms up to $\ell=20$
by the blue, purple and red solid curves, respectively.
It is clear to see that the Richardson transform indeed accelerates the convergence. 
We also show, in the right figure, how the Richardson transform $\cR_n$ with fixed $\ell_\text{max}=180$ works as $n$ grows.
In the current case, one should choose $n$ in the range $20 \lesssim n \lesssim 60$.
In the following analysis, we use the $30$th Richardson
transform.
Applying the Richardson transform to the sequence $a_\ell^{(1)}$, we get
\be
\cR_{30} [a_{150}^{(1)}]\approx -0.304297119449870831867025905573,
\ee
which is in agreement with \eqref{eq:A0} with 27-digit accuracy! 

Once $\mathcal{A}_1$ is fixed, we can proceed to the next-to-leading coefficients $\Gamma_1^{(1)}$.
We consider a sequence
\be
b_\ell^{(1)}=\( \ell -\frac{3}{2}\) \left[ \frac{\Gamma_\ell^{(0)}}{\mathcal{A}_1 \Gamma(\ell-\frac{1}{2})}-1 \right]
\ee
This should converge to $\Gamma_1^{(1)}$ in $\ell \to \infty$.
After the Richardson transform again, we find
\be
\cR_{30} [b_{150}^{(1)}]\approx -0.144860385419958982062924086.
\ee
The exact value of $\Gamma_1^{(1)}$ is found in \cite{BK2}:
\be
\Gamma_1^{(1)}=\frac{3(1-2\log 2)}{8}=-0.144860385419958982062924091\dots.
\ee
Both precisely coincide as expected.
Pushing the same computation, we can systematically confirm the relation \eqref{eq:LOB-pert} order by order.
In table~\ref{tab:LOB-pert}, we summarize the results estimated by the Richardson extrapolation.
The large order behavior indeed captures the information on the leading non-perturbative sector.

\begin{table}[tb]
\caption{The numerical estimation of $\Gamma_\ell^{(1)}$ from the large order behavior (LOB) in the perturbative expansion.
We use $\Gamma_\ell^{(0)}$ up to $\ell=180$ and the $30$th Richardson extrapolation.
We also show the deviation from the true $\Gamma_\ell^{(1)}$.}
\label{tab:LOB-pert}
\begin{center}
  \begin{tabular}{clc}\hline
$\ell$ & Estimation from LOB & $|1-\cR_{30}[\cdots]/\Gamma_\ell^{(1)}|$ \\ \hline  
$1$ & $-0.1448603854199589820629240863$ & $3.3 \times 10^{-26}$  \\
$2$ & $-0.0938694159959333439731081$ & $5.5 \times 10^{-23}$  \\
$3$ & $-0.14973248501597359266824$ & $5.3 \times 10^{-21}$  \\
$4$ & $-0.402728449526514898734$ & $2.2 \times 10^{-19}$  \\
$5$ & $-1.4747555976090392157$ & $5.9 \times 10^{-18}$  \\
$6$ & $-6.86483498151787623$ & $1.0 \times 10^{-16}$ \\ 
$7$ & $-38.674657843660018$ & $1.3 \times 10^{-15}$ \\ 
$8$ & $-255.9115123079947$ & $9.8 \times 10^{-15}$ \\ 
$9$ & $-1945.413350672867$ & $1.7 \times 10^{-14}$ \\ 
$10$ & $-16710.570082722$ & $ 9.3 \times 10^{-13}$ \\ 
\hline
\end{tabular}
\end{center}
\end{table}

Next, we want to test the second line in \eqref{eq:LOB-pert}.
To do so, we have to subtract the contribution in the first line.
This can be done by using the Borel resummation.
We first rewrite the first line in \eqref{eq:LOB-pert} as
\be
\ba
h^{(1)}(\ell):=1+\frac{\Gamma_1^{(1)}}{\ell-\frac{3}{2}}
+\frac{\Gamma_2^{(1)}}{(\ell-\frac{3}{2})(\ell-\frac{5}{2})}
+\cdots
=1+\sum_{m=1}^\infty \frac{\gamma_m^{(1)}}{\ell^m}
\ea
\label{eq:1inst-part}
\ee
The important point is that the coefficient $\gamma_m^{(1)}$ grows
factorially, and thus $h^{(1)}(\ell)$ is a formal divergent series.
Furthermore, we observe that the Borel transform $\cB[h^{(1)}](\zeta)$
has a singularity at $\zeta=\log 2$.
Thus $h^{(1)}(\ell)$ is non-Borel summable, and one has to use the lateral Borel resummation.
The ambiguity of the two lateral Borel resummations is roughly estimated as
\be
(\cS_+-\cS_-)h^{(1)}(\ell) \sim \ri \times \cO( 2^{-\ell}) ,
\label{eq:Sh-disc}
\ee
where $\cS_\pm=\cS_{0^\pm}$.
In other words, the imaginary parts of $\cS_\pm h^{(1)}(\ell)$ are of order $\cO( 2^{-\ell})$.
Now we consider a quantity
\be
\delta \Gamma_\ell^{(0)}:=\Gamma_\ell^{(0)}-\mathcal{A}_1 \Gamma\( \ell-\frac{1}{2} \) \cS_{+} h^{(1)}(\ell).
\ee
As mentioned above, since the imaginary parts of  $\cS_\pm h^{(1)}(\ell)$ behave as $\cO( 2^{-\ell})$,
$\delta \Gamma_\ell^{(0)}$ behaves as $\ri \times \cO(2^{-\ell})$ in $\ell \to \infty$.
This just corresponds to the contribution in the second line of \eqref{eq:LOB-pert}.%
\footnote{The relation \eqref{eq:LOB-pert} can be regarded as a ``transseries'' in the large $\ell$ expansion.
The second line is the leading ``non-perturbative'' correction in $1/\ell$, whose ``instanton action'' is $\log 2$.
The ambiguity of the lateral Borel resummations of the perturbative part (the first line) is almost
canceled by the ambiguity in the second line.
In other words, we can guess the contribution in the second line from that in the first line.}
Let us consider a sequence
\be
a_\ell^{(2)}=\frac{2^{\ell-1}\delta \Gamma_\ell^{(0)}}{\Gamma(\ell-1)}.
\ee
This should converge to $\mathcal{A}_2$.
As before, using the Richardson extrapolation, we find
\be
\cR_{20}[a_{160}^{(2)}]\approx 0.1454506142043354998\ri +3.0 \times 10^{-20},
\ee
whose imaginary part is in agreement with the exact $\mathcal{A}_2$ up to 17-digit accuracy.
As in the leading non-perturbative sector, one can estimate the coefficients
$\Gamma_\ell^{(2)}$ in \eqref{eq:LOB-pert}.
The obtained values up to $\ell=6$ are shown in table \ref{tab:LOB-pert2}.

\begin{table}[tb]
\caption{The numerical estimation of $\Gamma_\ell^{(2)}$ from the large order behavior of $\Gamma_\ell^{(0)}$.
We use $\Gamma_\ell^{(0)}$ up to $\ell=180$ and the $20$th Richardson extrapolation.}
\label{tab:LOB-pert2}
\begin{center}
\begin{tabular}{clc}\hline
$\ell$ & Estimation from LOB & $|1-\cR_{20}[\cdots]/\Gamma_\ell^{(2)}|$ \\ \hline  
$1$ & $-0.7499999999999987$ & $1.7 \times 10^{-15}$  \\
$2$ & $0.845472324053663$ & $7.0 \times 10^{-14}$  \\
$3$ & $-2.4877562953846$ & $8.8\times 10^{-13}$  \\
$4$ & $7.504881881284$ & $7.2 \times 10^{-12}$  \\
$5$ & $-37.76786450152$ & $1.1 \times 10^{-11}$  \\
$6$ & $186.98364073$ & $8.6 \times 10^{-10}$ \\ 
\hline
\end{tabular}
\end{center}
\end{table}

Finally, let us proceed to the third line in \eqref{eq:LOB-pert}.
At this level, it is not easy to extract the information with high precision by numerics.
We again consider quantities
\be
\ba
h^{(2)}(\ell):=1+\frac{2\Gamma_1^{(2)}}{\ell-2}+\frac{2^2 \Gamma_2^{(2)}}{(\ell-2)(\ell-3)}+\cdots
=1+\sum_{m=1}^\infty \frac{\gamma_m^{(2)}}{\ell^m}
\ea
\label{eq:2inst-part}
\ee
and
\be
\delta^2 \Gamma_\ell^{(0)}:=\Gamma_\ell^{(0)}-\mathcal{A}_1 \Gamma\( \ell-\frac{1}{2} \) \cS_{+} 
h^{(1)}(\ell)
-\mathcal{A}_2 \frac{\Gamma\( \ell-1 \)}{2^{\ell-1}} \cS_{+} h^{(2)}(\ell).
\ee
Defining two sequences%
\footnote{The reason of the separation of the sequence is purely technical. We observed that the Richardson extrapolation works better for this
rather than for the naive sequence.}
\be
\ba
a_\ell^{(3),\text{odd}}=\frac{3^{2\ell-1-\frac{3}{2}}}{\Gamma(2\ell-1-\frac{3}{2})}\delta^2 \Gamma_{2\ell-1}^{(0)}, \qquad
a_\ell^{(3),\text{even}}=\frac{3^{2\ell-\frac{3}{2}}}{\Gamma(2\ell-\frac{3}{2})}\delta^2 \Gamma_{2\ell}^{(0)},
\ea
\ee
we find
\be
\ba
\cR_{8}[a_{90}^{(3),\text{odd}}]&\approx0.101563618725,\\
\cR_{8}[a_{90}^{(3),\text{even}}]&\approx0.101563618713,
\ea
\label{eq:A3-LOB}
\ee
Unlike $\mathcal{A}_1$ and $\mathcal{A}_2$, these do not agree with the Stokes coefficient in the toy model:
\be
\frac{2 A^3}{\pi \ri}=0.0695237641771486153\dots .
\ee
This discrepancy is explained as follows.
As noted in section~\ref{sec:BES}, in the quantization condition \eqref{eq:QC2} up to order $\Lambda^4$, 
only the non-perturbative correction of $r(x_m)$ with $m=0$
gives a non-vanishing contribution.
However, at the order $\Lambda^6$, $r^\text{np}(x_m)$ with $m=1$ also contributes to the quantization condition.
As we will see in appendix~\ref{sec:3inst}, if taking into account this contribution, we find
the exact value of $\mathcal{A}_3$ as
\be
\ba
\mathcal{A}_3
=\frac{2A^3}{\pi \ri}\biggl(1+\frac{8\Gamma^4(\frac{5}{4})}{3\sqrt{3}\Gamma^4(\frac{3}{4})}\biggr) 
=0.101563618709385381\dots,
\ea
\ee
which is indeed very close to the numerical estimations \eqref{eq:A3-LOB} from the large order behavior!
This is precisely the expected value of the Stokes constant to reproduce the jump of the two transseries parameter responsible for the $\Lambda^6$ corrections, see the Appendix \ref{sec:3inst}, equation (\ref{eq:TSparam2}).
This result implies that at the order $\Lambda^{10}$, the Stokes coefficient also receives
a new contribution from $r^\text{np}(x_m)$ with $m=-1$.

The reason for this different behavior that begins only at the three instantons level $O(\Lambda^6)$ can be understood from the quantization condition.
The quantization condition (\ref{eq:QC2}) is a set of infinitely many linear equations in the undetermined functions $f_0(t)$ and $f_1(t)$. The coefficients appearing in these equations are the known functions $r(x_m)$ and at strong coupling they assume the transseries forms \eqref{eq:r-strongTS1}-\eqref{eq:r-strongTS2}. If we were to turn off all the non-perturbative corrections in all the $r(x_m)$, then our strong coupling perturbative expansion $f_0^{(0)}(t),f_1^{(0)}(t)$ in (\ref{eq:f0f1}), would solve this infinite set of equations.

We can proceed to turn on the first non-perturbative correction in the quantization condition (\ref{eq:QC2}).
As shown in table \ref{tab:r}, the first non-perturbative correction appears at order $\cO(\Lambda^2)$ in $r(x_0)$. This means that a non-perturbative correction of the same order has to be added to $f^{(0)}(t)$ and $f^{(1)}(t)$ and this correction has to be constructed entirely from the particular expansion \eqref{eq:r-strongTS1}-\eqref{eq:r-strongTS2} combined with the perturbative expansions $f_0^{(0)}(t),f_1^{(0)}(t)$, as in \cite{BK2}. The next non-perturbative correction is of order $\cO(\Lambda^4)$ and it is once again coming from the transseries expansion of $r(x_0)$. A correction of the same order will now be needed in $f^{(0)}(t)$ and $f^{(1)}(t)$ and this correction has to be constructed from the order $\cO(\Lambda^4)$ of $r(x_0)$ combined with our perturbative expansions $f_0^{(0)}(t),f_1^{(0)}(t)$, and from the order $\cO(\Lambda^2)$ of $r(x_0)$ combined with the $\cO(\Lambda^2)$ just obtained for $f_0^{(0)}(t),f_1^{(0)}(t)$.

From table \ref{tab:r}, we see that at order $\cO(\Lambda^6)$ we will receive two contributions, see \eqref{eq:r-strongTS1}-\eqref{eq:r-strongTS2}, one from $r(x_0)$ and one from $r(x_1)$. The correction of this order to $f^{(0)}(t)$ and $f^{(1)}(t)$ can still be obtained from the lower orders, but it stops being a one parameter transseries and becomes a two parameter transseries, see Appendix~\ref{sec:3inst}. Clearly, as we see from table \ref{tab:r}, at $\cO(\Lambda^{10})$ the non-perturbative corrections to $r(x_1)$ will come into play and we will need yet another transseries parameter and so on so forth.
  
The above large order behavior, as well as our analysis of the quantization condition equations, strongly indicates that the strong coupling expansion of the physical cusp anomalous dimension
is a \textit{multi-parameter transseries}, while the expansion in the toy model is
a one-parameter transseries.
This is a big difference between the toy model and the true cusp anomalous dimension. 

\paragraph{Large order behavior of the mass gap.}

Let us turn to the large order behavior of the mass gap coefficients $m_\ell^{(n)}$.
Using the results derived in the Appendix \ref{sec:3inst}, we computed the large order behaviors of the perturbative expansion
$m_\text{O(6)}^{(0)}$ in \eqref{eq:mass-trans}, as well as for the coefficients of the first two instantons corrections $m_\text{O(6)}^{(1)}$ and $m_\text{O(6)}^{(2)}$.

As expected from the discussion in Section \ref{sec:Borel}, the perturbative coefficients $m_\ell^{(0)}$ behave as
\begin{align}
m_\ell^{(0)}\sim &\notag\frac{A}{2\pi \ri} \Gamma\left(\ell-\frac{1}{2}\right) \left(1+ \frac{m_1^{(1)}}{\ell-\frac{3}{2} } +\cO(\ell^{-2})\right)+\frac{A^2}{2\pi \ri} \frac{\Gamma\left(\ell-1\right)}{2^{\ell-1}} \left(1+ \frac{2m_1^{(2)}}{\ell-2 } +\cO(\ell^{-2})\right)\\
&+\frac{B}{2\pi \ri} \frac{\Gamma\left(\ell-\frac{1}{2}\right)}{3^{\ell-1/2}} \left(1+\cO(\ell^{-1})\right)+...\,,
\end{align}
where the Stokes constant $A$ is precisely the one obtained in the toy model \eqref{eq:toyStokes} or in the cusp anomalous dimension.

The Stokes constant for the three instantons sector $B$ is once again not consistent with a one parameter transseries ansatz, i.e. it is not simply $B\neq A^3$, see also the discussion at the end of Section \ref{subsec:LOB}.
At the three instantons level a new transseries parameter arise and, from the correct multi parameter transseries form derived in the Appendix \eqref{eq:mO6TS}, we can deduce
\begin{equation}
B= -\frac{\ri\,\pi}{8 \sqrt{3} \Gamma(\frac{3}{4})^2}.
\end{equation}

For the large order coefficients in the first non-perturbative correction $m_\text{O(6)}^{(1)}$ in \eqref{eq:mass-trans}, we can easily recognize the presence of the perturbative coefficients as well as the higher instantons one:
\begin{align}
m_\ell^{(1)}\sim &\notag\frac{C}{2\pi \ri} \frac{\Gamma\left(\ell+\frac{1}{2}\right)}{(-1)^\ell} \left(1+ \frac{-m_1^{(0)}}{\ell-\frac{1}{2} } +\cO(\ell^{-2})\right)+\frac{2A}{2\pi \ri} \Gamma\left(\ell-\frac{1}{2}\right) \left(1+ \frac{m_1^{(2)}}{\ell-3/2 } +\cO(\ell^{-2})\right)\\
&+\frac{3A^2}{2\pi \ri} \frac{\Gamma\left(\ell-1\right)}{2^{\ell-1}} \left(1+\cO(\ell^{-1})\right)+...\,,
\end{align}
where the Stokes constant $C$ is associated with the presence of a Stokes line in the $\theta = \pi $ direction, with a characteristic alternating factor $1/(-2\pi)^\ell$ in front of it, and it is given by

\begin{equation}
C= \frac{2\pi\ri }{\Gamma\left(\frac{3}{4}\right)^2}\,.
\end{equation}

The presence of this Stokes line is obvious since the one instanton sector will ``see" the perturbative sector with a relative action $2\pi \times (-1)$, 
hence the alternating nature of the coefficients.
Note that this is precisely the same Stokes constant in the $\theta=\pi$ direction that one can obtain from the toy model!

\subsection{Lateral Borel resummations and ambiguity cancellation}
In this subsection, we perform the Borel resummation of the strong coupling expansion.
As emphasized in \cite{BKK}, the perturbative strong coupling expansion of $\Gc(g)$ is
a \textit{non-Borel summable asymptotic expansion}.
This is because that the Borel transform of $\Gc^{(0)}(g)$ has the singularities on the positive real axis
in the Borel plane, as shown in figure~\ref{fig:Borel-sing}.
One has to avoid these singularities when performing the inverse 
Borel transform (or Borel resummation).
There is an ambiguity to choose such integration contours, and the lateral Borel resummations are not
real-valued any more.
The imaginary part must be canceled by the other contributions, i.e., the non-perturbative sectors, in the full transseries expansion
\eqref{eq:cusp-trans} because the cusp anomalous dimension is a real-valued quantity for $g \in \mathbb{R}$.
Here we test this cancellation by explicitly performing the lateral Borel resummations of $\Gc^{(n)}(g)$ for $n=0,1,2$.

Let us consider the inverse Borel transform of \eqref{eq:Borel-Gamma}.
By using the well-known integral representation of the gamma function:
\be
\Gamma(z)=\int_0^\infty \rd \zeta \, \re^{-\zeta} \zeta^{z-1}, \qquad \real z>0,
\label{eq:Gamma-int-rep}
\ee
it is easy to see that the inverse Borel transform is given by
\be
\ba
\cS_0 \Gc^{(n)}(g)&=1+\int_0^\infty \rd \zeta \, \re^{-\zeta} \zeta^{-3/2} \widetilde{\cB}[ \Gc^{(n)} ]\( \frac{\zeta}{2\pi g} \) \\
&=1+ (2\pi g)^{-1/2} \int_0^\infty \rd \zeta\,  \re^{-2\pi g \zeta} \zeta^{-3/2} \widetilde{\cB}[ \Gc^{(n)} ](\zeta).
\ea
\ee
where we have separated the $\ell=0$ term from the others to use the integral representation \eqref{eq:Gamma-int-rep}.
This resummation is, however, not well-defined because $\cB[ \Gc^{(n)} ](\zeta)$
has singularities on the positive real axis.
Therefore we need to modify it as the lateral Borel resummations:
\be
\cS_{\pm}  \Gc^{(n)}(g)=1+ (2\pi g)^{-1/2} \int_{0}^{\infty \re^{\pm \ri 0}}  
\rd \zeta\,  \re^{-2\pi g \zeta} \zeta^{-3/2} \widetilde{\cB}[ \Gc^{(n)} ](\zeta).
\label{eq:lateralBorel}
\ee
These lateral Borel resummations are complex-valued quantities, and
cause the discontinuity $(\cS_{+} -\cS_{-} ) \Gc^{(n)}(g)$.

Now we perform the lateral Borel-Pad\'e resummations at finite coupling.
Let us see in detail for $g=1/5$ as an example. 
Using the formula \eqref{eq:lateralBorel}, we find the following numerical values
\be
\ba
\cS_{\pm}  \Gc^{(0)}(1/5)&\approx 0.1252604 \mp 0.2097335 \ri, \\
\cS_{\pm}  \Gc^{(1)}(1/5)&\approx 0.8678150 \mp 0.0811188 \ri, \\
\cS_{\pm}  \Gc^{(2)}(1/5)&\approx 0.6406029 \mp 0.0846409 \ri.
\ea
\ee
These are complex numbers, and obviously the imaginary part in the perturbative part is not negligible.
More importantly, these resummations take the different values depending on the integration contours.
Therefore there is an ambiguity of the resummations.
Plugging these values into the transseries \eqref{eq:cusp-trans} 
and setting the parameter as $\sigma=\re^{\mp 3\pi \ri/4}$ \cite{BK2} for
$\cS_{\pm}$, respectively, we finally get
\begin{align}
&\cS_{\pm}  \Gc^{(0)}-\frac{\Lambda_\pm^2}{2\pi g}\cS_{\pm}  \Gc^{(1)}
+\frac{\Lambda_\pm^4}{16\pi^2 g^2}\cS_{\pm}  \Gc^{(2)} \biggr|_{g=\frac{1}{5}} \notag \\
&\approx (0.1252604 \mp 0.2097335 \ri)+(0.2303182 \pm 0.1909411 \ri)+(0.00249307 \pm 0.01886876 \ri) \notag \\
&\approx 0.3580716 \pm 0.0000763 \ri, 
\end{align}
where $\Lambda_\pm^2$ are defined by \eqref{eq:Lambda-pm}.
The imaginary part becomes very small after taking into account the non-perturbative corrections!
Furthermore, the real part is close to the true value directly evaluated from the
BES equation in table~\ref{tab:cusp-num}.
We also show the ambiguity cancellation for $g=1/2,1,2$ in tables~\ref{tab:BorelResum1}, \ref{tab:BorelResum2} 
and \ref{tab:BorelResum3}.
The real parts are in remarkable agreement with the values in table~\ref{tab:cusp-num}.
We conclude that the ambiguity of the lateral Borel resummations are precisely canceled by
the perturbative and non-perturbative contributions
and that the finial results are unambiguous real values.

\begin{table}[tb]
\caption{The ambiguity cancellation in the lateral Borel resummations at $g=1/2$. The total imaginary part is indeed 
very close to zero due to the cancellation.
The real part should be compared with the directly evaluated values by the method in section~\ref{sec:BES}.}
\label{tab:BorelResum1}
\begin{center}
\begin{tabular}{cl}\hline
Contributions & \hspace{1.5cm}Values \\ \hline 
$\cS_{\pm} \Gc^{(0)}$ & $0.63182936 \mp 0.02188690 \ri$  \\
$-\frac{\Lambda_\pm^2}{2\pi g}\cS_{\pm}  \Gc^{(1)}$ & $0.02210593 \pm  0.02166913 \ri$   \\
$\frac{\Lambda_\pm^4}{16\pi^2 g^2}\cS_{\pm}  \Gc^{(2)}$ & $0.00000326\pm 0.00021822 \ri$  \\
\hline
Sum & $0.65393855 \pm 0.00000044\ri$ \\
\hline
Direct Evaluation & $0.65393903$ \\
\hline
\end{tabular}
\end{center}
\end{table}

\begin{table}[tb]
\caption{The ambiguity cancellation at $g=1$.}
\label{tab:BorelResum2}
\begin{center}
\begin{tabular}{cl}\hline
Contributions & \hspace{2.2cm}Values \\ \hline 
$\cS_{\pm} \Gc^{(0)}$ & $0.825965638690 \mp 0.000693266968 \ri$  \\
$-\frac{\Lambda_\pm^2}{2\pi g}\cS_{\pm}  \Gc^{(1)}$ & $0.000693493818 \pm 0.000693040133 \ri$   \\
$\frac{\Lambda_\pm^4}{16\pi^2 g^2}\cS_{\pm}  \Gc^{(2)}$ & $0.000000000111\pm 0.000000226842\ri$  \\
\hline
Sum & $0.826659132619 \pm 0.000000000007\ri$ \\
\hline
Direct Evaluation & $0.826659132627$ \\
\hline
\end{tabular}
\end{center}
\end{table}

\begin{table}[tb]
\caption{The ambiguity cancellation at $g=2$.}
\label{tab:BorelResum3}
\begin{center}
\begin{tabular}{cl}\hline
Contributions & \hspace{4cm}Values \\ \hline 
$\cS_{\pm} \Gc^{(0)}$ & $0.915571204186344585360 \mp 0.000000928959331963843 \ri$  \\
$-\frac{\Lambda_\pm^2}{2\pi g}\cS_{\pm}  \Gc^{(1)}$ & 
$0.000000928959749702527 \pm 0.000000928958914225168 \ri$   \\
$\frac{\Lambda_\pm^4}{16\pi^2 g^2}\cS_{\pm}  \Gc^{(2)}$ & 
$0.000000000000000000282 \pm 0.000000000000417738679\ri$  \\
\hline
Sum & $0.915572133146094288169 \pm 0.000000000000000000005\ri$ \\
\hline
Direct Evaluation & $0.915572133146094288174$ \\
\hline
\end{tabular}
\end{center}
\end{table}

\subsection{Novel relations}
In this subsection, we observe quite novel relations between the cusp anomalous dimension and the mass gap in the O(6) sigma model.
So far, we do not have a derivation of these relations from the BES equation.
It would be important to prove them along the line \cite{BK2}.
We first find that the strong coupling expansion of the mass gap \eqref{eq:mass-trans} has the following beautiful relation
\be
\frac{ m_\text{O(6)}^{(2)}}{m_\text{O(6)}^{(1)}}=\frac{ m_\text{O(6)}^{(1)}}{m_\text{O(6)}^{(0)}}
=1-\frac{0.6051396146}{2\pi g}+\frac{0.8516809822}{(2\pi g)^2}-\frac{2.271453077}{(2\pi g)^3}+\cdots.
\ee
We checked this equality up to $1/g^{180}$ with sufficient numerical precision.
Also, we find the similar relation to \eqref{eq:cusp-mass-1}:
\be
\sum_{\ell=0}^\infty \frac{\Gamma_\ell^{(2)}}{(2\pi g)^{\ell}}
=\left[ \sum_{\ell=0}^\infty \frac{m_{\ell}^{(0)}}{(2\pi g)^{\ell}} \right]\left[ \sum_{\ell=0}^\infty \frac{m_{\ell}^{(1)}}{(2\pi g)^{\ell}} \right].
\label{eq:cusp-mass-2}
\ee
These relations suggest us the following guess:
\be
\ba
\frac{ m_\text{O(6)}^{(n)}}{m_\text{O(6)}^{(n-1)}} \stackrel{?}{=} \frac{ m_\text{O(6)}^{(1)}}{m_\text{O(6)}^{(0)}}, \qquad\quad
\sum_{\ell=0}^\infty \frac{\Gamma_\ell^{(n)}}{(2\pi g)^{\ell}}
\stackrel{?}{=} \left[ \sum_{\ell=0}^\infty \frac{m_{\ell}^{(0)}}{(2\pi g)^{\ell}} \right]\left[ \sum_{\ell=0}^\infty \frac{m_{\ell}^{(n-1)}}{(2\pi g)^{\ell}} \right].
\ea
\label{eq:guess}
\ee
In the toy model in the previous section, we have already seen that the similar relations hold for all $n$.
However, in the physical case, this is probably not the case for $n=3$.
A reason of this is very likely due to the contribution from $r^\text{np}(x_1)$.%
\footnote{In other words, if we turn off this correction by hand: $ r^\text{np}(x_1)=0$, then the relations \eqref{eq:guess}
seem to work for $n=3$. Of course, this does not give the true cusp anomalous dimension.}
This contribution should change the structure of the coefficients $\Gamma_\ell^{(3)}$, but the right hand side in \eqref{eq:guess}
there is no source of such a change.
It would be interesting to check it in more detail by computing the higher non-perturbative corrections
from the BES equation.

\section{Conclusions and future directions}\label{sec:conc}

The cusp anomalous dimension has a rich non-perturbative structure at strong coupling.
We can put it on the resurgent analysis, recently developed in many contexts.
Solving the BES equation at strong coupling systematically,
we computed the perturbative and non-perturbative corrections up to very high orders
in the $1/g$ expansions.
Using these data, we showed that the large order behavior in the perturbative expansion has the non-perturbative
information, as expected.
Moreover, the ambiguity of the Borel resummations in the perturbative sector is precisely canceled by
the contributions from the non-perturbative sectors.
The final result is real-valued and in remarkable agreement with the direct evaluation at finite (not so strong) coupling.
We also find the unexpected relations between the cusp anomalous dimension and the mass gap of the O(6) sigma model.

In this paper, we focused on the cusp anomalous dimension.
Clearly, it would be interesting to explore the strong coupling expansions for other examples. 
Closely related examples are the generalized scaling function, proposed in \cite{FRS}, 
and the generalized cusp anomalous dimension (or equivalently the quark-antiquark potential),
studied in \cite{DF, CHMS1, CHMS2, Drukker, CMS}.
As studied in \cite{BK1, FGR1, FGR2, FGR3}, the strong coupling analysis of the generalized scaling function 
is almost in parallel with the cusp anomalous dimension,
and thus it is a good exercise to see its resurgent aspect along the line in this paper.
The analysis of the generalized cusp anomalous dimension will be much more involved.

Another example is the eigenvalue of the adjoint BFKL operator.
This eigenvalue plays a very important role in scattering amplitudes in the so-called multi-Regge limit.
Recently, Basso, Caron-Huot and Sever proposed integral equations that compute the adjoint BFKL eigenvalue
at any coupling \cite{BCHS} (see also \cite{Hatsuda:2014oza}). 
Since these equations are quite similar to the BES equation, it seems to be possible to use the technique in section~\ref{sec:BES}.
It would be interesting to ask whether the BFKL eigenvalue receives non-perturbative corrections 
at strong coupling or not.

The strong coupling limit of short operators is also an important problem.
In particular, the Konishi operator is a significant example.
In \cite{GKV2, Frolov}, 
it was observed from the TBA analysis \cite{GKV1, Bombardelli:2009ns, Gromov:2009bc, Arutyunov:2009ur} 
that the conformal dimension of the Konishi operator has the strong coupling expansion
\be
\Delta_\text{K}-2=2\sqrt[4]{\lambda}\(1+\frac{1}{\sqrt{\lambda}}+\cdots \), \qquad
\lambda \to \infty.
\ee
The same result was also reproduced from the various worldsheet computations \cite{Roiban:2009aa, Gromov:2011de, Roiban:2011fe, Vallilo:2011fj}.
It would be important to clarify the analytic structure of the Konishi operator at strong coupling.
The Quantum Spectral Curve formulation \cite{GKLV1, GKLV2} will be helpful for this purpose.
See \cite{Gromov:2014bva} for an interesting approach to the strong coupling Konishi dimension, based on \cite{Basso:2011rs}.

The generalization to the ABJM theory \cite{ABJM} is also interesting.
The all-loop Bethe ansatz equation in planar ABJM theory was conjectured
by Gromov and Vieira in \cite{Gromov:2008qe}.
One important consequence is that in the integrability approach, the 't Hooft coupling
always appears through a non-trivial function $h(\lambda)$, called the interpolating function.
Then, it turned out in \cite{Gromov:2008qe} that the cusp anomalous dimension in ABJM theory is related to that in $\cN=4$ SYM by
\be
\Gc^\text{ABJM}(\lambda)=\frac{1}{2} \Gc^{\cN=4}(g=h(\lambda)).
\ee
It was a long-standing open problem to determine $h(\lambda)$.
Surprisingly, Gromov and Sizov  recently proposed an exact formula of $h(\lambda)$ \cite{Gromov:2014eha}.
According to their result, the interpolating function at strong coupling behaves as
\be
h(\lambda)=\sqrt{\frac{1}{2}\( \lambda-\frac{1}{24} \)}-\frac{\log 2}{2\pi}+\cO(\re^{-2\pi\sqrt{2\lambda}}),
\qquad \lambda \to \infty.
\label{eq:h-strong}
\ee 
Now it is clear to see that $\Gc^\text{ABJM}(\lambda)$ receives two kinds of non-perturbative corrections at strong coupling.
One is just obtained by replacing $g$ in $\Gc^{\cN=4}(g)$ by $h(\lambda)$:
\be
\cO(\re^{-2\pi h(\lambda)}) \sim \cO(\re^{-\pi \sqrt{2\lambda}}).
\label{eq:former-np}
\ee
The other is the correction%
\footnote{This non-perturbative correction is quite different from the first type correction \eqref{eq:former-np}
even though the exponential factors are almost same.
The strong coupling perturbative expansion of $h(\lambda)$ in \eqref{eq:h-strong} is a \textit{convergent} series.
Thus one does not need the Borel resummation.
Nevertheless, $h(\lambda)$ receives non-perturbative corrections!
This non-perturbative structure is essentially the same as that in the $1/2$ BPS Wilson loop \cite{Drukker:2009hy}, 
analyzed in \cite{MP1, DMP1}.
In fact, the idea in \cite{Gromov:2014eha} is to relate the interpolating function to the results in \cite{MP1, DMP1}.
} 
that $h(\lambda)$ itself receives, as in \eqref{eq:h-strong},
\be
\cO(\re^{-2\pi \sqrt{2\lambda}}).
\ee
It would be interesting to discuss physical origins of these two corrections in the effective worldsheet theory on $AdS_4 \times \mathbb{CP}^3$
in the high spin limit.
As discussed in \cite{DMP1}, the latter type correction is probably related to stringy worldsheet instantons, constructed in \cite{Cagnazzo:2009zh}.

\acknowledgments{We thank Benjamin Basso, Misha Isachenkov, Marcos Mari\~no, Kazumi Okuyama,  and Marcel Vonk for valuable discussions.}

\appendix

\section{Special functions}\label{sec:special}

Let us introduce some functions used in the main text. See appendix~D in \cite{BK2} in more detail.
We first consider the following series:
\be
F(a,b|z) := \sum_{k=0}^\infty \frac{(a)_k (b)_k }{(1)_k} z^k.\label{eq:F-def}
\ee
where we used the Pochhammer symbol $(x)_n=\Gamma(x+n)/\Gamma(x)$.
This series is asymptotic. 
Let us consider the following ``Borel transform'':
\be
\widehat{\cB} F(\zeta) =\sum_{k=0}^\infty \frac{(b)_k}{(1)_k}\frac{\zeta^k}{k!}
=(1-\zeta)^{-b}.
\label{eq:Borel-F}
\ee
This Borel transform has a branch cut along $(1, \infty)$ if $b$ is a non-integer.
Therefore for $z<0$, $F(a,b|z)$ is Borel summable.
Since the series is non-Borel summable for $z>0$, one needs the lateral Borel resummations.
One easily finds that the inverse Borel transform of \eqref{eq:Borel-F} is given by
\be
S_{0^\pm} F(a,b|z)=\frac{1}{\Gamma(a)} \int_0^{\infty \re^{\pm \ri 0}} 
\rd \zeta \, \re^{-\zeta} \zeta^{a-1} (1-z \zeta)^{-b}.
\ee
A simple computation shows that the discontinuity of these lateral Borel resummations is given by
\be
(\cS_{0^+}-\cS_{0^-})F(a,b|z)=2\pi \ri \frac{z^{1-a-b}}{\Gamma(a)\Gamma(b)} \re^{-1/z} \cS_0 F(1-a,{1-b}| {-z}),
\qquad z>0.
\label{eq:disc}
\ee
Remarkably, the Borel resummations satisfy the following Wronskian-like relation \cite{BK2}
\be
\cS_{0^\pm} F \( \frac{1}{4},\frac{1}{4} \bigg| {z} \) \cS_0 F\( -\frac{1}{4},\frac{3}{4} \bigg| {-z} \)
-\frac{z}{4} \cS_{0^\pm} F \( \frac{1}{4},\frac{5}{4}\bigg| {z} \) \cS_0 F\( \frac{3}{4},\frac{3}{4} \bigg| {-z}\)=1,\label{eq:Wronskian}
\ee
where we assumed $z>0$.

Next, let us define the functions $V_n(x)$ and $U_n^\pm(x)$ ($n=0,1$) by
\be\ba
V_n(x) &:= \frac{\sqrt{2}}{\pi} \int_{-1}^1 \rd u \, (1+u)^{1/4-n}(1-u)^{-1/4} \re^{u x}, \\
U_n^\pm(x) &:= \frac{1}{2} \int_1^\infty \rd u\, (u \pm 1)^{-1/4} (u \mp 1)^{1/4-n} \re^{-(u-1)x}, \quad (\real x >0).
\ea
\label{eq:Vn-def}
\ee
Note that the integral in $V_n(x)$ is well-defined for all complex values of $x$, but the integrals in $U_n^\pm(x)$
are convergent only for $\real x>0$.
For $\real x<0$, we need analytic continuations of $U_n^\pm(x)$.
As mentioned in \cite{BK2}, the function $V_n(x)$ is written in terms of the Whittaker function of the first kind
\be
V_n(x)=\frac{1}{2^n} \frac{\Gamma(\frac{5}{4}-n)}{\Gamma(\frac{5}{4})\Gamma(2-n)} (-2x)^{n/2-1} M_{1/4-n/2,1/2-n/2}(-2x).
\label{eq:Vn-M}
\ee
Also, the functions $U_n^\pm(x)$ are related to the Whittaker function of the second kind:
\be
\ba
U_0^+(x)&=\frac{1}{2} \Gamma\( \frac{5}{4} \) x^{-1} \re^x W_{-1/4,1/2}(2x), \\
U_0^-(x)&=\frac{1}{2} \Gamma\( \frac{3}{4} \) x^{-1} \re^x W_{1/4,1/2}(2x),
\ea
\qquad
\ba
U_1^+(x)&=\frac{1}{2} \Gamma\( \frac{1}{4} \) (2x)^{-1/2} \re^x W_{1/4,0}(2x), \\
U_1^-(x)&=\frac{1}{2} \Gamma\( \frac{3}{4} \) (2x)^{-1/2} \re^x W_{-1/4,0}(2x).
\ea
\ee
These relations naturally give analytic continuations for $\real x < 0$.
Note that these functions have branch cuts on the negative axis.
As shown in \cite{BK2}, the functions $V_n(x)$ and $U_n^\pm(x)$ are related by
\be
\ba
V_0(x)&=\frac{2\sqrt{2}}{\pi} \re^{\mp 3\pi \ri/4} [ \re^x U_0^-(-x)+\re^{-x} U_0^+(x) ], \\
V_1(x)&=\frac{2\sqrt{2}}{\pi} \re^{\mp 3\pi \ri/4} [ \re^x U_1^-(-x)-\re^{-x} U_1^+(x) ],
\ea
\label{eq:V-U}  
\ee
where the upper and lower signs correspond to $\im x >0$ and $\im x<0$, respectively.
For real $x$, one needs the $\ri \epsilon$-prescription on the right hand sides due to the branch cut of $U_n^\pm(x)$.

Now we rewrite $U_n^\pm(x)$ in terms of the Borel resummation of $F(a,b|z)$.
As in \cite{BK2}, for $x>0$ we have
\be
\ba
U_0^+(x)&=(2x)^{-5/4}\Gamma\( \frac{5}{4} \) \cS_0 F\( \frac{1}{4},\frac{5}{4} \Big| {-\frac{1}{2x}} \) ,\\
U_0^-(x)&=(2x)^{-3/4}\Gamma\( \frac{3}{4} \) \cS_0 F\( -\frac{1}{4},\frac{3}{4} \Big| {-\frac{1}{2x}} \)  ,\\
U_1^+(x)&=\frac{1}{2} (2x)^{-1/4}\Gamma\( \frac{1}{4} \) \cS_0 F\( \frac{1}{4},\frac{1}{4} \Big| {-\frac{1}{2x}} \), \\
U_1^-(x)&=\frac{1}{2} (2x)^{-3/4}\Gamma\( \frac{3}{4} \) \cS_0 F\( \frac{3}{4},\frac{3}{4} \Big| {-\frac{1}{2x}} \).
\ea
\qquad\quad  (x >0),
\ee
As mentioned before, $F(a,b|z)$ is Borel summable for $z<0$.
If the argument of $U_n^\pm(x)$ is negative, we need the $\ri \epsilon$-presciption.
It just corresponds to the lateral Borel resummations of $F(a,b|z)$ for $z>0$:
\be
\ba
U_0^+(-x \pm \ri \epsilon )&=\re^{\pm 3\pi \ri/4} (2x)^{-5/4}\Gamma\( \frac{5}{4} \) \cS_{0^\pm} F\( \frac{1}{4},\frac{5}{4} \Big| {\frac{1}{2x}} \) ,\\
U_0^-(-x \pm \ri \epsilon)&=\re^{\mp 3\pi \ri/4}(2x)^{-3/4}\Gamma\( \frac{3}{4} \) \cS_{0^\pm}  F\( -\frac{1}{4},\frac{3}{4} \Big| {\frac{1}{2x}} \)  ,\\
U_1^+(-x \pm \ri \epsilon)&=-\frac{1}{2}\re^{\pm 3\pi \ri/4} (2x)^{-1/4}\Gamma\( \frac{1}{4} \) \cS_{0^\pm}  F\( \frac{1}{4},\frac{1}{4} \Big| {\frac{1}{2x}} \), \\
U_1^-(-x \pm \ri \epsilon)&=\frac{1}{2}\re^{\mp 3\pi \ri/4} (2x)^{-3/4}\Gamma\( \frac{3}{4} \) \cS_{0^\pm}  F\( \frac{3}{4},\frac{3}{4} \Big| {\frac{1}{2x}} \).
\ea
\qquad\quad  (x > 0),
\ee
Substituting these expressions into \eqref{eq:V-U}, one can express $V_n(x)$ in terms of $F(a,b|z)$.
In particular, one finds
\be
\ba
V_0(-\pi g)&=\frac{(2\pi g)^{-5/4} \re^{\pi g}}{\Gamma(\frac{3}{4})}
\left[ \cS_{0^\pm} F\( \frac{1}{4},\frac{5}{4} \Big| {\frac{1}{2\pi g}} \)+\Lambda_\pm^2  \cS_0 F\( -\frac{1}{4},\frac{3}{4} \Big| {-\frac{1}{2\pi g}} \) \right], \\
V_1(-\pi g)&=\frac{(2\pi g)^{-5/4} \re^{\pi g}}{2\Gamma(\frac{3}{4})}
\left[ 8\pi g\, \cS_{0^\pm} F\( \frac{1}{4},\frac{1}{4} \Big| {\frac{1}{2\pi g}} \)+\Lambda_\pm^2  \cS_0 F\( \frac{3}{4},\frac{3}{4} \Big| {-\frac{1}{2\pi g}} \) \right],
\ea
\label{eq:V0V1}
\ee
where the non-perturbative scales $\Lambda_\pm$ are defined by \eqref{eq:Lambda} with $\sigma=\sigma_\pm=\re^{\mp 3\pi \ri/4}$,
respectively.
Note that these expressions for upper and lower signs are equivalent because of the discontinuity \eqref{eq:disc} of the lateral Borel resummations.
In other words, the ambiguity of the lateral Borel resummations is precisely canceled by the non-perturbative correction (the second term in \eqref{eq:V0V1}).

It is useful to write down the expression \eqref{eq:r} with $x=x_m=m-\frac{1}{4}$ at strong coupling.
As noted before, the strong coupling behavior is different for $m \geq 1$ and $m\leq 0$.
For $m \geq 1$, we obtain
\be
\ba
r(x_m)=\frac{1}{2}\frac{F(\frac{3}{4},\frac{3}{4}|\frac{\alpha}{4m-1})-(-1)^m 4(4m-1)^{\frac{1}{2}}\(\frac{\Gamma(\frac{5}{4})}{\Gamma(\frac{3}{4})}\)^{4m}
\alpha^{2m-1}\Lambda^{8m-2}F(\frac{1}{4},\frac{1}{4}|{-\frac{\alpha}{4m-1}})}
{F(-\frac{1}{4},\frac{3}{4}|\frac{\alpha}{4m-1})+(-1)^m (4m-1)^{-\frac{1}{2}}\(\frac{\Gamma(\frac{5}{4})}{\Gamma(\frac{3}{4})}\)^{4m}
\alpha^{2m}\Lambda^{8m-2}F(\frac{1}{4},\frac{5}{4}|{-\frac{\alpha}{4m-1}})},
\ea
\label{eq:r-strong-1}
\ee
where $\alpha=\frac{1}{2\pi g}$.
Similarly, we find the expression for $m \leq 0$,
\be
\ba
r(x_m)
&=\frac{2(1-4m)}{\alpha}\\
& \times\frac{F(\frac{1}{4},\frac{1}{4}|\frac{\alpha}{1-4m})+(-1)^m \frac{1}{4}(1-4m)^{-\frac{1}{2}}\(\frac{\Gamma(\frac{5}{4})}{\Gamma(\frac{3}{4})}\)^{-4m}
\alpha^{-2m+1}\Lambda^{2-8m}F(\frac{3}{4},\frac{3}{4}|{-\frac{\alpha}{1-4m}})}
{F(\frac{1}{4},\frac{5}{4}|\frac{\alpha}{1-4m})+(-1)^m (1-4m)^{\frac{1}{2}}\(\frac{\Gamma(\frac{5}{4})}{\Gamma(\frac{3}{4})}\)^{-4m}
\alpha^{-2m}\Lambda^{2-8m}F(-\frac{1}{4},\frac{3}{4}|{-\frac{\alpha}{1-4m}})},
\ea
\label{eq:r-strong-2}
\ee
It is obvious to see that the non-perturbative correction to $r(x_m)$ starts from $\cO(\Lambda^{|8m-2|})$.

With the above expression (\ref{eq:r-strong-2}) for $r(x_m)$ we can easily obtain its transseries expansion.
Let us assume that $\mbox{Re} \,\alpha >0 $, then for $m>0$ we can write
\begin{align}
r(x_m) = &\notag\frac{1}{2}\frac{F\left(\frac{3}{4},\frac{3}{4}| \frac{\alpha}{(4m-1)} \right)}{F\left(-\frac{1}{4},\frac{3}{4}| \frac{\alpha}{(4m-1)}  
\right)}
+2\sum_{n=1}^\infty (-1)^{(m+1)n} \left(\frac{\Gamma(5/4)}{\Gamma(3/4)}\right)^{4mn}  
 (4m-1)^{1-n/2}\\
 &\label{eq:r-strongTS1}\Lambda_{\pm}^{2(4m-1)n} \alpha^{2mn-1} \frac{F\left(\frac{1}{4},\frac{5}{4}| -\frac{\alpha}{(4m-1)} \right)^{n-1}}{F\left(-\frac{1}{4},\frac{3}{4}| \frac{\alpha}{(4m-1)}  \right)^{n+1}}\,,
\end{align}
where the choice on transseries parameter $\sigma_{\pm}$ depends on $\mbox{Im}\, \alpha >0 $ or $\mbox{Im}\, \alpha<0$ (note that for $\alpha= 1/(2\pi g)$ this corresponds to $\mbox{Im}\, g <0 $, or $\mbox{Im}\, g >0 $ respectively).
Similarly for $-m\leq0$, assuming always $\mbox{Re}\, \alpha >0 $, we have
\begin{align}
r(x_{-m}) = &\notag\frac{2(4m+1)}{\alpha} \frac{ F\left( \frac{1}{4},\frac{1}{4}| \frac{\alpha}{(4m+1)} \right) }{ F\left(\frac{1}{4},\frac{5}{4}|  \frac{\alpha}{(4m+1)} \right)}+2\sum_{n=1}^\infty (-1)^{(m+1)n} \left(\frac{\Gamma(5/4)}{\Gamma(3/4)}\right)^{4mn}  
 (4m+1)^{1+n/2}\\
 &\label{eq:r-strongTS2}\Lambda_{\pm}^{2(4m+1)n}\,\alpha^{2mn-1} \frac{F\left(-\frac{1}{4},\frac{3}{4}| - \frac{\alpha}{(4m+1)} \right)^{n-1} }{F\left(-\frac{1}{4},\frac{5}{4}|  \frac{\alpha}{(4m+1)}\right)^{n+1} }\,,
\end{align}
where once again the choice of transseries parameter $\sigma_{\pm}$ inside $\Lambda_{\pm}$ is correlated with $\mbox{Im}\, \alpha>0$ or $\mbox{Im}\, \alpha<0$.

\section{Strong coupling expansions}\label{sec:strong}

Here we explicitly give the strong coupling expansion up to the next-to-leading non-perturbative sector.%
\footnote{More precise and higher order coefficients are available on request to the authors.}
The perturbative strong coupling expansion is 
\be
\ba
\Gc^{(0)}(g) &= 1-\frac{1.039720771}{2\pi g}-\frac{0.2289913985}{(2\pi g)^2}-\frac{0.3648665524}{(2\pi g)^3}-\frac{0.9405461014}{(2\pi g)^4} \\
&\quad-\frac{3.356917310}{(2\pi g)^5}-\frac{15.29423354}{(2\pi g)^6}-\frac{84.82236515}{(2\pi g)^7}-\frac{554.5319782}{(2\pi g)^8}\\
&\quad-\frac{4176.215725}{(2\pi g)^9}-\frac{35606.52529}{(2\pi g)^{10}}+\cO(g^{-11}).
\ea
\ee
These are, of course, in agreement with the result in \cite{BKK}.
The expansion around the leading non-perturbative sector is also given by
\be
\ba
\Gc^{(1)}(g) &= 1-\frac{0.1448603854}{2\pi g}-\frac{0.09386941600}{(2\pi g)^2}-\frac{0.1497324850}{(2\pi g)^3}-\frac{0.4027284495}{(2\pi g)^4} \\
&\quad-\frac{1.474755598}{(2\pi g)^5}-\frac{6.864834982}{(2\pi g)^6}-\frac{38.67465784}{(2\pi g)^7}-\frac{255.9115123}{(2\pi g)^8}\\
&\quad-\frac{1945.413351}{(2\pi g)^9}-\frac{16710.57008}{(2\pi g)^{10}}+\cO(g^{-11}).
\ea
\ee
The expansion of $\Gc^{(2)}(g)$ is
\be
\ba
\Gc^{(2)}(g) &= 1-\frac{0.7500000000}{2\pi g}+\frac{0.8454723241}{(2\pi g)^2}-\frac{2.487756295}{(2\pi g)^3}+\frac{7.504881881}{(2\pi g)^4} \\
&\quad-\frac{37.76786450}{(2\pi g)^5}+\frac{186.9836409}{(2\pi g)^6}-\frac{1324.269347}{(2\pi g)^7}+\frac{9157.551684}{(2\pi g)^8}\\
&\quad-\frac{83418.23010}{(2\pi g)^9}+\frac{741922.2110}{(2\pi g)^{10}}+\cO(g^{-11}).
\ea
\ee
Similarly, the ``perturbative'' part of the mass gap in \eqref{eq:mass-trans} is
\be
\ba
m_\text{O(6)}^{(0)}&= 1-\frac{0.07243019271}{2\pi g}-\frac{0.04955777441}{(2\pi g)^2}-\frac{0.07845572166}{(2\pi g)^3}-\frac{0.2082747743}{(2\pi g)^4}\\
&\quad-\frac{0.7563512718}{(2\pi g)^5}-\frac{3.500599444}{(2\pi g)^6}-\frac{19.64470145}{(2\pi g)^7}-\frac{129.6331369}{(2\pi g)^8}\\
&\quad-\frac{983.5017470}{(2\pi g)^9}-\frac{8435.500952}{(2\pi g)^{10}}+\cO(g^{-11}).
\ea
\ee
The expansions in the non-perturbative sectors are given by
\be
\ba
m_\text{O(6)}^{(1)}&= 1-\frac{0.6775698073}{2\pi g}+\frac{0.8459535867}{(2\pi g)^2}-\frac{2.381606844}{(2\pi g)^3}+\frac{7.529420762}{(2\pi g)^4}\\
&\quad-\frac{36.65893377}{(2\pi g)^5}+\frac{187.6790274}{(2\pi g)^6}-\frac{1294.485118}{(2\pi g)^7}+\frac{9189.267411}{(2\pi g)^8}\\
&\quad-\frac{81900.06954}{(2\pi g)^9}+\frac{744213.7198}{(2\pi g)^{10}}+\cO(g^{-11}).
\ea
\ee
and
\be
\ba
m_\text{O(6)}^{(2)}&= 1-\frac{1.282709422}{2\pi g}+\frac{2.107658901}{(2\pi g)^2}-\frac{5.742053268}{(2\pi g)^3}+\frac{18.79808052}{(2\pi g)^4}\\
&\quad-\frac{85.81920322}{(2\pi g)^5}+\frac{440.6817456}{(2\pi g)^6}-\frac{2896.496146}{(2\pi g)^7}+\frac{20598.28223}{(2\pi g)^8}\\
&\quad-\frac{177671.1890}{(2\pi g)^9}+\frac{1.620648725\times 10^6}{(2\pi g)^{10}}+\cO(g^{-11}).
\ea
\ee

\section{Three instantons corrections}\label{sec:3inst}

As we mentioned in the main text, the solution to the quantization condition \eqref{eq:QC2}, and hence the cusp anomalous dimension, stop being a one parameter transseries at the order $\cO(\Lambda^6)$.
To show this we analyzed the solution to \eqref{eq:QC2} using the methods of \cite{BK2}.
As shown in \cite{BK2}, we can rewrite \eqref{eq:QC2} as
\be
\ba
&\sum_{n \geq 1} c_+(n,g) \frac{n U_0^+(4\pi n g) r(x_m)+U_1^+(4\pi n g)x_m}{n-x_m}\\
&+\sum_{n \geq 1} c_-(n,g) \frac{n U_0^-(4\pi n g) r(x_m)+U_1^-(4\pi n g)x_m}{n+x_m}=1, 
\ea\label{eq:QCc}
\ee
with $m\in\mathbb{Z}$, where the undetermined functions $f_0(t)$ and $f_1(t)$ are written in terms of the coefficients 
$c_\pm(n,g) $
\begin{align}
f_0(t) &= -1 +\sum_{n=1}^\infty t\left(c_+ (n,g) \frac{U_1^+(4\pi n g)}{4\pi n g -t}+c_- (n,g) \frac{U_1^-(4\pi n g)}{4\pi n g +t}\right)\,,\\
f_1(t) &= \sum_{n=1}^\infty 4\pi n g \left(c_+ (n,g) \frac{U_0^+(4\pi n g)}{4\pi n g -t}+c_- (n,g) \frac{U_1^-(4\pi n g)}{4\pi n g +t}\right)\,.
\end{align}
These coefficients can be found by using the transseries ansatz
\be
c_\pm(n,g)=(8\pi n g)^{\pm 1/4} \sum_{m=0}^\infty \Lambda^{2m} \sum_{k=0}^\infty \frac{c_\pm^{(m,k)}(n)}{(4\pi g)^k},\qquad
\Lambda \sim \cO(g^{1/2}\re^{-2\pi g})\,,
\ee
and solving order by order in $1/g$ and order by order in $\Lambda$ the infinite set of equations \eqref{eq:QCc}.

Then, proceeding in the same way as \cite{BK2}, we find the first few coefficients.
In the perturbative sector, the solutions are
\be
\ba
&
\ba
c_+^{(0,0)}(n)&=a_+(n) \equiv \frac{2\Gamma(n+\frac{1}{4})}{\Gamma^2 (\frac{1}{4}) \Gamma(n+1)},\\
c_+^{(0,1)}(n)&=-a_+(n)\(\frac{3\log 2}{4}+\frac{3}{32n} \),
\ea
\qquad
\ba
c_-^{(0,0)}(n)&=a_-(n) \equiv \frac{\Gamma(n+\frac{3}{4})}{2\Gamma^2 (\frac{3}{4}) \Gamma(n+1)},\\
c_-^{(0,1)}(n)&=a_-(n)\(\frac{3\log 2}{4}+\frac{5}{32n} \),
\ea \\
&\, c_+^{(0,2)}(n)= a_+(n)\( \frac{\mathrm{K}}{8}-\frac{27 \log^2 2}{32}-\frac{27\log 2}{128n}-\frac{63}{2048n^2} \), \\
&\, c_-^{(0,2)}(n)= a_-(n)\( \frac{7\mathrm{K}}{8}+\frac{45 \log^2 2}{32}+\frac{75\log 2}{128n}+\frac{225}{2048n^2} \),
\ea
\ee
In the leading non-perturbative sector, we find
\be
\ba
&
\ba
c_+^{(1,0)}(n)&=0,\\
c_+^{(1,1)}(n)&=-\frac{1}{2}a_+(n),
\ea
\qquad
\ba
c_-^{(1,0)}(n)&=a_-(n-1),\\
c_-^{(1,1)}(n)&=a_-(n-1) \( \frac{1}{4} -\frac{3\log 2}{4} +\frac{1}{32n} \),
\ea \\
&\, c_+^{(1,2)}(n)=-a_+(n) \( -\frac{3}{8}+\frac{3\log 2}{8}+\frac{9}{64n} \) , \\
&\, c_-^{(1,2)}(n)=a_-(n-1) \biggl(  -\frac{15}{32}+\frac{\mathrm{K}}{8}+\frac{9\log 2}{16}-\frac{27\log^2 2}{32} 
+\frac{-7+9\log 2}{128n}+\frac{25}{2048n^2} \biggr).
\ea
\ee
At the next-to-leading order, we get
\be
\ba
c_+^{(2,0)}(n)&=0,\\
c_+^{(2,1)}(n)&=0,\\
c_+^{(2,2)}(n)&=\frac{1}{4}a_+(n),
\ea
\qquad
\ba
c_-^{(2,0)}(n)&=0,\\
c_-^{(2,1)}(n)&=-\frac{1}{2}a_-(n-1),\\
c_-^{(2,2)}(n)&=a_-(n-1)\(1-\frac{3 \log 2}{8}-\frac{1}{64n} \).
\ea
\ee

Up to this order all of the above coefficients are perfectly consistent with a one parameter transseries expansion for the cusp anomalous dimension, essentially because up to this order we could have truncated to the perturbative level all but the $m=0$ equations in \eqref{eq:QCc}.
At the three instanton level though, the transseries expansion \eqref{eq:r-strongTS1} for $r(x_1)$ contains a $\Lambda^6$ term.
\be
\ba
c_+^{(3,0)}(n)&=0,\\
c_+^{(3,1)}(n)&=0,\\
c_+^{(3,2)}(n)&=-a_+(n-1)\(4 \frac{\Gamma(5/4)^4}{\sqrt{3} \Gamma(3/4)^4}\),
\ea
\qquad
\ba
c_-^{(3,0)}(n)&=0,\\
c_-^{(3,1)}(n)&=0,\\
c_-^{(3,2)}(n)&=\frac{1}{4}a_-(n-1).
\ea
\ee
Using all of these results, we get the following strong coupling expansion of $\Gc(g)$:
\be
\ba
\frac{\Gc(g)}{2g}=\Gc^{(0)}(g)-\frac{\Lambda^2}{2\pi g} \Gc^{(1)}(g)+\frac{ \Lambda^4}{16\pi^2 g^2} \Gc^{(2)}(g)
-\frac{\Lambda^6}{128\pi^3 g^3}\Gc^{(3)}(g)+\cO(\Lambda^8),
\ea
\ee
where
\be
\ba
\Gc^{(0)}(g)&=1-\frac{3 \log 2}{4\pi g}-\frac{\mathrm{K}}{16\pi^2 g^2}+\( -\frac{ 3 \log 2}{64\pi^3} \mathrm{K}-\frac{27}{2048\pi^3} \)\frac{1}{g^3}+\cO(g^{-4}),\\
\Gc^{(1)}(g)&=1+\frac{3(1-2\log 2)}{16\pi g}+ \frac{8\mathrm{K}-9(3-4\log 2+4\log^2 2)}{512\pi^2 g^2} +\cO(g^{-3}),\\
\Gc^{(2)}(g)&=1-\frac{3}{8\pi g}+\cO(g^{-2}),\\
\Gc^{(3)}(g)&=1 - 16 \frac{\Gamma(\frac{5}{4})^4}{3 \sqrt{3} \Gamma(\frac{3}{4})^4}+\cO(g^{-1})\,.
\ea
\label{eq:Gamma-asym}
\ee

These values are all consistent with the results presented in Appendix \ref{sec:strong}.

Note that the leading order of $\Gc^{(3)}(g)$ is different from one, this means that the order $\cO(\Lambda^6)$ should actually be rewritten as 
\begin{equation}
\frac{\Lambda^6}{128\pi^3 g^3}\Gc^{(3)}(g)= \frac{\sigma_{\pm}^3+{\tilde{\sigma}}_{\pm}}{128 \pi^3 g^3} \left(\sqrt{2\pi g} \re^{-2\pi g}\right)^3\,,
\end{equation}
with the usual transseries parameter $\sigma_{\pm}=\re^{\mp3\pi \ri/4} \Gamma(\frac{3}{4})/\Gamma(\frac{5}{4})$ and the new second transseries parameter 
\begin{equation}
\tilde{\sigma}_{\pm}=-\re^{\mp3\pi \ri/4} \frac{16 \Gamma(\frac{5}{4})}{3 \sqrt{3} \Gamma(\frac{3}{4})}\,.
\label{eq:TSparam2}
\end{equation}

From the correct transseries expansion for the coefficients $c_{\pm}(n,g)$, we can easily obtain the mass gap transseries expansion from equation (\ref{eq:massgap}).
As it turns out also the mass gap is described by an transseries with infinitely many transseries parameters and at order $\Lambda^6$ is given by
\begin{align}
m_\text{O(6)}=\frac{\sqrt{2}}{\Gamma(\frac{5}{4})}(2\pi g)^{1/4} \re^{-\pi g}&\notag \left[ \left(1 +\frac{3-6\log 2}{32\pi g}+\frac{16K-63+198 \log2-108(\log 2)^2}{2048(\pi g)^2}+O(g^{-3})\right)\right.\\
&\notag - \frac{\Lambda_{\pm}^2}{8 \pi g}\left(1 -\frac{15-6\log 2}{32\pi g}+O(g^{-2})\right) +\frac{\Lambda_{\pm}^4}{(8 \pi g)^2}(1 +O(g^{-1}))\\
&\label{eq:mO6TS}\left.+ \frac{\Lambda_{\pm}^6}{(8 \pi g)^2} \,\frac{4 \sqrt{3}}{3} (1 +O(g^{-1})\right]\,.
\end{align}
Note precisely how the $\Lambda^6$ leading order is not the one that could be guessed from a one parameter transseries ansatz!

Obviously also these values are all consistent with the results presented in Appendix \ref{sec:strong}.


\begin{thebibliography}{00}

\bibitem{Maldacena} 
  J.~M.~Maldacena,
  ``The Large N limit of superconformal field theories and supergravity,''
  Adv.\ Theor.\ Math.\ Phys.\  {\bf 2}, 231 (1998)
  [hep-th/9711200].

\bibitem{GKP1} 
  S.~S.~Gubser, I.~R.~Klebanov and A.~M.~Polyakov,
  ``Gauge theory correlators from noncritical string theory,''
  Phys.\ Lett.\ B {\bf 428}, 105 (1998)
  [hep-th/9802109].
  
  
\bibitem{Witten} 
  E.~Witten,
  ``Anti-de Sitter space and holography,''
  Adv.\ Theor.\ Math.\ Phys.\  {\bf 2}, 253 (1998)
  [hep-th/9802150].
  
  
\bibitem{Integrability} 
  N.~Beisert, C.~Ahn, L.~F.~Alday, Z.~Bajnok, J.~M.~Drummond, L.~Freyhult, N.~Gromov and R.~A.~Janik {\it et al.},
  ``Review of AdS/CFT Integrability: An Overview,''
  Lett.\ Math.\ Phys.\  {\bf 99}, 3 (2012)
  [arXiv:1012.3982 [hep-th]].

\bibitem{GKLV1} 
  N.~Gromov, V.~Kazakov, S.~Leurent and D.~Volin,
  ``Quantum Spectral Curve for Planar $\mathcal{N} = 4$ Super-Yang-Mills Theory,''
  Phys.\ Rev.\ Lett.\  {\bf 112}, no. 1, 011602 (2014)
  [arXiv:1305.1939 [hep-th]].

\bibitem{GKLV2} 
  N.~Gromov, V.~Kazakov, S.~Leurent and D.~Volin,
  ``Quantum spectral curve for arbitrary state/operator in AdS$_5$/CFT$_4$,''
  arXiv:1405.4857 [hep-th].

\bibitem{AM1} 
  L.~F.~Alday and J.~M.~Maldacena,
  ``Gluon scattering amplitudes at strong coupling,''
  JHEP {\bf 0706}, 064 (2007)
  [arXiv:0705.0303 [hep-th]].

\bibitem{Drummond:2007aua} 
  J.~M.~Drummond, G.~P.~Korchemsky and E.~Sokatchev,
  ``Conformal properties of four-gluon planar amplitudes and Wilson loops,''
  Nucl.\ Phys.\ B {\bf 795}, 385 (2008)
  [arXiv:0707.0243 [hep-th]].

\bibitem{Brandhuber:2007yx} 
  A.~Brandhuber, P.~Heslop and G.~Travaglini,
  ``MHV amplitudes in N=4 super Yang-Mills and Wilson loops,''
  Nucl.\ Phys.\ B {\bf 794}, 231 (2008)
  [arXiv:0707.1153 [hep-th]].
  
\bibitem{Drummond:2007au} 
  J.~M.~Drummond, J.~Henn, G.~P.~Korchemsky and E.~Sokatchev,
  ``Conformal Ward identities for Wilson loops and a test of the duality with gluon amplitudes,''
  Nucl.\ Phys.\ B {\bf 826}, 337 (2010)
  [arXiv:0712.1223 [hep-th]].

\bibitem{BES} 
  N.~Beisert, B.~Eden and M.~Staudacher,
  ``Transcendentality and Crossing,''
  J.\ Stat.\ Mech.\  {\bf 0701}, P01021 (2007)
  [hep-th/0610251].

\bibitem{ES} 
  B.~Eden and M.~Staudacher,
  ``Integrability and transcendentality,''
  J.\ Stat.\ Mech.\  {\bf 0611}, P11014 (2006)
  [hep-th/0603157].
  

 \bibitem{ABJM}
 O.~Aharony, O.~Bergman, D.~L.~Jafferis and J.~Maldacena, ``N=6 superconformal Chern-Simons-matter theories, M2-branes and their gravity duals,''
  JHEP {\bf 0810}, 091 (2008)
  [arXiv:0806.1218 [hep-th]].

\bibitem{DMP1} 
  N.~Drukker, M.~Marino and P.~Putrov,
  ``From weak to strong coupling in ABJM theory,''
  Commun.\ Math.\ Phys.\  {\bf 306}, 511 (2011)
  [arXiv:1007.3837 [hep-th]].

\bibitem{Gromov:2014eha} 
  N.~Gromov and G.~Sizov,
  ``Exact Slope and Interpolating Functions in N=6 Supersymmetric Chern-Simons Theory,''
  Phys.\ Rev.\ Lett.\  {\bf 113}, no. 12, 121601 (2014)
  [arXiv:1403.1894 [hep-th]].

\bibitem{Bern:2006ew} 
  Z.~Bern, M.~Czakon, L.~J.~Dixon, D.~A.~Kosower and V.~A.~Smirnov,
  ``The Four-Loop Planar Amplitude and Cusp Anomalous Dimension in Maximally Supersymmetric Yang-Mills Theory,''
  Phys.\ Rev.\ D {\bf 75}, 085010 (2007)
  [hep-th/0610248].

\bibitem{Cachazo:2006az} 
  F.~Cachazo, M.~Spradlin and A.~Volovich,
  ``Four-loop cusp anomalous dimension from obstructions,''
  Phys.\ Rev.\ D {\bf 75}, 105011 (2007)
  [hep-th/0612309].

\bibitem{BBKS} 
  M.~K.~Benna, S.~Benvenuti, I.~R.~Klebanov and A.~Scardicchio,
  ``A Test of the AdS/CFT correspondence using high-spin operators,''
  Phys.\ Rev.\ Lett.\  {\bf 98}, 131603 (2007)
  [hep-th/0611135].

\bibitem{AABEK} 
  L.~F.~Alday, G.~Arutyunov, M.~K.~Benna, B.~Eden and I.~R.~Klebanov,
  ``On the Strong Coupling Scaling Dimension of High Spin Operators,''
  JHEP {\bf 0704}, 082 (2007)
  [hep-th/0702028 [HEP-TH]].

\bibitem{Kostov:2007kx} 
  I.~Kostov, D.~Serban and D.~Volin,
  ``Strong coupling limit of Bethe ansatz equations,''
  Nucl.\ Phys.\ B {\bf 789}, 413 (2008)
  [hep-th/0703031 [HEP-TH]].

\bibitem{Beccaria:2007tk} 
  M.~Beccaria, G.~F.~De Angelis and V.~Forini,
  ``The Scaling function at strong coupling from the quantum string Bethe equations,''
  JHEP {\bf 0704}, 066 (2007)
  [hep-th/0703131].

\bibitem{BKK} 
  B.~Basso, G.~P.~Korchemsky and J.~Kotanski,
  ``Cusp anomalous dimension in maximally supersymmetric Yang-Mills theory at strong coupling,''
  Phys.\ Rev.\ Lett.\  {\bf 100}, 091601 (2008)
  [arXiv:0708.3933 [hep-th]].


\bibitem{GKP2} 
  S.~S.~Gubser, I.~R.~Klebanov and A.~M.~Polyakov,
  ``A Semiclassical limit of the gauge / string correspondence,''
  Nucl.\ Phys.\ B {\bf 636}, 99 (2002)
  [hep-th/0204051].

\bibitem{Kruczenski:2007cy} 
  M.~Kruczenski, R.~Roiban, A.~Tirziu and A.~A.~Tseytlin,
  ``Strong-coupling expansion of cusp anomaly and gluon amplitudes from quantum open strings in $AdS_5 \times S^5$,''
  Nucl.\ Phys.\ B {\bf 791}, 93 (2008)
  [arXiv:0707.4254 [hep-th]].

\bibitem{Roiban:2007jf} 
  R.~Roiban, A.~Tirziu and A.~A.~Tseytlin,
  ``Two-loop world-sheet corrections in $AdS_5 \times S^5$ superstring,''
  JHEP {\bf 0707}, 056 (2007)
  [arXiv:0704.3638 [hep-th]].

\bibitem{Roiban:2007dq} 
  R.~Roiban and A.~A.~Tseytlin,
  ``Strong-coupling expansion of cusp anomaly from quantum superstring,''
  JHEP {\bf 0711}, 016 (2007)
  [arXiv:0709.0681 [hep-th]].



  \bibitem{Ecalle:1981}
J.~Ecalle, `` Les Fonctions Resurgentes,'' vol.~I - III.
\newblock Publ. Math. Orsay, 1981.

\bibitem{ZinnJustin:1980uk}
J.~Zinn-Justin, ``Perturbation Series at Large Orders in Quantum Mechanics
  and Field Theories: Application to the Problem of Resummation,''  Phys. Rept.
  {\bf 70} (1981) 109.

\bibitem{Voros:1983}
A.~Voros, ``The Return of the Quartic Oscillator: The Complex WKB Method,''
   Ann. Inst. Henri Poincar{\'e} {\bf 39} (1983) 211.


\bibitem{Dunne:2012zk}
G.~V. Dunne and M.~{\"U}nsal,``Continuity and Resurgence: Towards a
  Continuum Definition of the ${\mathbb{CP}}^{N-1}$ Model,''  Phys. Rev. {\bf D87}
  (2013) 025015.

\bibitem{Argyres:2012vv}
P.~Argyres and M.~{\"U}nsal, ``A Semiclassical Realization of Infrared
  Renormalons,''  Phys. Rev. Lett. {\bf 109} (2012) 121601.

\bibitem{Cherman:2013yfa}
A.~Cherman, D.~Dorigoni, G.~V. Dunne and M.~{\"U}nsal, ``Resurgence in QFT:
  Unitons, Fractons and Renormalons in the Principal Chiral Model,''
  Phys.\ Rev.\ Lett.\  {\bf 112} (2014) 021601.


\bibitem{BK1} 
  B.~Basso and G.~P.~Korchemsky,
  ``Embedding nonlinear O(6) sigma model into N=4 super-Yang-Mills theory,''
  Nucl.\ Phys.\ B {\bf 807}, 397 (2009)
  [arXiv:0805.4194 [hep-th]].

\bibitem{BBBKP} 
  Z.~Bajnok, J.~Balog, B.~Basso, G.~P.~Korchemsky and L.~Palla,
  ``Scaling function in AdS/CFT from the O(6) sigma model,''
  Nucl.\ Phys.\ B {\bf 811}, 438 (2009)
  [arXiv:0809.4952 [hep-th]].


\bibitem{BK2} 
  B.~Basso and G.~P.~Korchemsky,
  ``Nonperturbative scales in AdS/CFT,''
  J.\ Phys.\ A {\bf 42}, 254005 (2009)
  [arXiv:0901.4945 [hep-th]].

\bibitem{AM2} 
  L.~F.~Alday and J.~M.~Maldacena,
  ``Comments on operators with large spin,''
  JHEP {\bf 0711}, 019 (2007)
  [arXiv:0708.0672 [hep-th]].

\bibitem{Basso:2010in} 
  B.~Basso,
  ``Exciting the GKP string at any coupling,''
  Nucl.\ Phys.\ B {\bf 857}, 254 (2012)
  [arXiv:1010.5237 [hep-th]].
  
 
\bibitem{Dunne:2015ywa}
  G.~V.~Dunne and M.~Unsal,
  ``Resurgence and Dynamics of O(N) and Grassmannian Sigma Models,''
  arXiv:1505.07803 [hep-th].
  
\bibitem{Dorigoni:2014hea} 
  D.~Dorigoni,
  ``An Introduction to Resurgence, transseries and Alien Calculus,''
  arXiv:1411.3585 [hep-th].
  
  

\bibitem{Aniceto:2015rua}
  I.~Aniceto,
  ``The Resurgence of the Cusp Anomalous Dimension,''
  arXiv:1506.03388 [hep-th].
  

\bibitem{Aniceto:2011nu}
I.~Aniceto, R.~Schiappa and M.~Vonk, ``The Resurgence of Instantons in
  String Theory,''   Commun. Num. Theor. Phys. {\bf 6} (2012) 339.


\bibitem{berry1993unfolding}
	M.~V.~Berry and C.~J.~Howls, ``Unfolding the high orders of asymptotic expansions with coalescing saddles: singularity theory, crossover and duality,'' 
	 Proc. R. Soc. A {\bf{443}} (1993), 107-126.
  
\bibitem{BalianPV}
R.~Balian, G.~Parisi, and A.~Voros, ``Discrepancies from Asymptotic Series and Their Relation to Complex Classical Trajectories,'' 
Phys. Rev. Lett. {\bf 41}, (1978) 1141; Erratum Phys. Rev. Lett. {\bf 41}, (1978) 1627.

\bibitem{Couso-Santamaria:2014iia}
  R.~C.~Santamaria, J.~D.~Edelstein, R.~Schiappa and M.~Vonk,
  ``Resurgent Transseries and the Holomorphic Anomaly: Nonperturbative Closed Strings in Local CP2,'' 
  arXiv:1407.4821 [hep-th].

\bibitem{Vonk:2015sia}
  M.~Vonk, ``Resurgence and Topological Strings,'' 
  arXiv:1502.05711 [hep-th].
  


\bibitem{Kostov:2008ax} 
  I.~Kostov, D.~Serban and D.~Volin,
  ``Functional BES equation,''
  JHEP {\bf 0808}, 101 (2008)
  [arXiv:0801.2542 [hep-th]].

\bibitem{Volin} 
  D.~Volin,
  ``Quantum integrability and functional equations: Applications to the spectral problem of AdS/CFT and two-dimensional sigma models,''
  J.\ Phys.\ A {\bf 44}, 124003 (2011)
  [arXiv:1003.4725 [hep-th]].


  
\bibitem{Marino:2007te} 
  M.~Marino, R.~Schiappa and M.~Weiss,
  ``Nonperturbative Effects and the Large-Order Behavior of Matrix Models and Topological Strings,''
  Commun.\ Num.\ Theor.\ Phys.\  {\bf 2}, 349 (2008)
  [arXiv:0711.1954 [hep-th]].


\bibitem{FRS} 
  L.~Freyhult, A.~Rej and M.~Staudacher,
  ``A Generalized Scaling Function for AdS/CFT,''
  J.\ Stat.\ Mech.\  {\bf 0807}, P07015 (2008)
  [arXiv:0712.2743 [hep-th]].

\bibitem{DF} 
  N.~Drukker and V.~Forini,
  ``Generalized quark-antiquark potential at weak and strong coupling,''
  JHEP {\bf 1106}, 131 (2011)
  [arXiv:1105.5144 [hep-th]].

\bibitem{CHMS1} 
  D.~Correa, J.~Henn, J.~Maldacena and A.~Sever,
  ``An exact formula for the radiation of a moving quark in N=4 super Yang Mills,''
  JHEP {\bf 1206}, 048 (2012)
  [arXiv:1202.4455 [hep-th]].

\bibitem{CHMS2} 
  D.~Correa, J.~Henn, J.~Maldacena and A.~Sever,
  ``The cusp anomalous dimension at three loops and beyond,''
  JHEP {\bf 1205}, 098 (2012)
  [arXiv:1203.1019 [hep-th]].

\bibitem{Drukker} 
  N.~Drukker,
  ``Integrable Wilson loops,''
  JHEP {\bf 1310}, 135 (2013)
  [arXiv:1203.1617 [hep-th]].

\bibitem{CMS} 
  D.~Correa, J.~Maldacena and A.~Sever,
  ``The quark anti-quark potential and the cusp anomalous dimension from a TBA equation,''
  JHEP {\bf 1208}, 134 (2012)
  [arXiv:1203.1913 [hep-th]].

\bibitem{FGR1} 
  D.~Fioravanti, P.~Grinza and M.~Rossi,
  ``Strong coupling for planar N=4 SYM theory: An All-order result,''
  Nucl.\ Phys.\ B {\bf 810}, 563 (2009)
  [arXiv:0804.2893 [hep-th]].

\bibitem{FGR2} 
  D.~Fioravanti, P.~Grinza and M.~Rossi,
  ``The Generalised scaling function: A Note,''
  Nucl.\ Phys.\ B {\bf 827}, 359 (2010)
  [arXiv:0805.4407 [hep-th]].

\bibitem{FGR3} 
  D.~Fioravanti, P.~Grinza and M.~Rossi,
  ``The Generalised scaling function: A Systematic study,''
  JHEP {\bf 0911}, 037 (2009)
  [arXiv:0808.1886 [hep-th]].

\bibitem{BCHS} 
  B.~Basso, S.~Caron-Huot and A.~Sever,
  ``Adjoint BFKL at finite coupling: a short-cut from the collinear limit,''
  JHEP {\bf 1501}, 027 (2015)
  [arXiv:1407.3766 [hep-th]].



\bibitem{Hatsuda:2014oza} 
  Y.~Hatsuda,
  ``Wilson loop OPE, analytic continuation and multi-Regge limit,''
  JHEP {\bf 1410}, 38 (2014)
  [arXiv:1404.6506 [hep-th]].


\bibitem{GKV2} 
  N.~Gromov, V.~Kazakov and P.~Vieira,
  ``Exact Spectrum of Planar ${\cal N}=4$ Supersymmetric Yang-Mills Theory: Konishi Dimension at Any Coupling,''
  Phys.\ Rev.\ Lett.\  {\bf 104}, 211601 (2010)
  [arXiv:0906.4240 [hep-th]].

\bibitem{Frolov} 
  S.~Frolov,
  ``Konishi operator at intermediate coupling,''
  J.\ Phys.\ A {\bf 44}, 065401 (2011)
  [arXiv:1006.5032 [hep-th]].



\bibitem{GKV1} 
  N.~Gromov, V.~Kazakov and P.~Vieira,
  ``Exact Spectrum of Anomalous Dimensions of Planar N=4 Supersymmetric Yang-Mills Theory,''
  Phys.\ Rev.\ Lett.\  {\bf 103}, 131601 (2009)
  [arXiv:0901.3753 [hep-th]].

\bibitem{Bombardelli:2009ns} 
  D.~Bombardelli, D.~Fioravanti and R.~Tateo,
  ``Thermodynamic Bethe Ansatz for planar AdS/CFT: A Proposal,''
  J.\ Phys.\ A {\bf 42}, 375401 (2009)
  [arXiv:0902.3930 [hep-th]].

\bibitem{Gromov:2009bc} 
  N.~Gromov, V.~Kazakov, A.~Kozak and P.~Vieira,
  ``Exact Spectrum of Anomalous Dimensions of Planar N = 4 Supersymmetric Yang-Mills Theory: TBA and excited states,''
  Lett.\ Math.\ Phys.\  {\bf 91}, 265 (2010)
  [arXiv:0902.4458 [hep-th]].

\bibitem{Arutyunov:2009ur} 
  G.~Arutyunov and S.~Frolov,
  ``Thermodynamic Bethe Ansatz for the AdS(5) x S(5) Mirror Model,''
  JHEP {\bf 0905}, 068 (2009)
  [arXiv:0903.0141 [hep-th]].

\bibitem{Roiban:2009aa} 
  R.~Roiban and A.~A.~Tseytlin,
  ``Quantum strings in $AdS_5 \times S^5$: Strong-coupling corrections to dimension of Konishi operator,''
  JHEP {\bf 0911}, 013 (2009)
  [arXiv:0906.4294 [hep-th]].

\bibitem{Gromov:2011de} 
  N.~Gromov, D.~Serban, I.~Shenderovich and D.~Volin,
  ``Quantum folded string and integrability: From finite size effects to Konishi dimension,''
  JHEP {\bf 1108}, 046 (2011)
  [arXiv:1102.1040 [hep-th]].

\bibitem{Roiban:2011fe} 
  R.~Roiban and A.~A.~Tseytlin,
  ``Semiclassical string computation of strong-coupling corrections to dimensions of operators in Konishi multiplet,''
  Nucl.\ Phys.\ B {\bf 848}, 251 (2011)
  [arXiv:1102.1209 [hep-th]].

\bibitem{Vallilo:2011fj} 
  B.~C.~Vallilo and L.~Mazzucato,
  ``The Konishi multiplet at strong coupling,''
  JHEP {\bf 1112}, 029 (2011)
  [arXiv:1102.1219 [hep-th]].

\bibitem{Gromov:2014bva} 
  N.~Gromov, F.~Levkovich-Maslyuk, G.~Sizov and S.~Valatka,
  ``Quantum spectral curve at work: from small spin to strong coupling in $ \mathcal{N} $ = 4 SYM,''
  JHEP {\bf 1407}, 156 (2014)
  [arXiv:1402.0871 [hep-th]].

\bibitem{Basso:2011rs} 
  B.~Basso,
  ``An exact slope for AdS/CFT,''
  arXiv:1109.3154 [hep-th].


\bibitem{Gromov:2008qe} 
  N.~Gromov and P.~Vieira,
  ``The all loop AdS4/CFT3 Bethe ansatz,''
  JHEP {\bf 0901}, 016 (2009)
  [arXiv:0807.0777 [hep-th]].


\bibitem{Drukker:2009hy} 
  N.~Drukker and D.~Trancanelli,
  ``A Supermatrix model for N=6 super Chern-Simons-matter theory,''
  JHEP {\bf 1002}, 058 (2010)
  [arXiv:0912.3006 [hep-th]].

\bibitem{MP1} 
  M.~Marino and P.~Putrov,
  ``Exact Results in ABJM Theory from Topological Strings,''
  JHEP {\bf 1006}, 011 (2010)
  [arXiv:0912.3074 [hep-th]].

\bibitem{Cagnazzo:2009zh} 
  A.~Cagnazzo, D.~Sorokin and L.~Wulff,
  ``String instanton in AdS(4) x CP**3,''
  JHEP {\bf 1005}, 009 (2010)
  [arXiv:0911.5228 [hep-th]].


\end{thebibliography}
\end{document}